\documentclass[twocolumn,letterpaper,aps,prc,superscriptaddress,showpacs,nofootinbib,floatfix]{revtex4-1}
\usepackage{multirow}
\usepackage{rotating}
\usepackage{notes2bib}
\usepackage{longtable}  
\usepackage{graphicx}
\usepackage{amsmath,amssymb,bbold,bm}
\usepackage{epstopdf}
\usepackage{xcolor}
\usepackage{lipsum}
\usepackage{hyperref}
\usepackage{url}
\usepackage{xcolor}

\def\nuc#1#2{${}^{#1}$#2}

\def\nonubb{$0\nu\beta\beta$}
\def\twonubb{$2\nu\beta\beta$}

\def\MJ{{\sc Majorana}}

\def\MJD{{\sc Majorana Demonstrator}}

\begin{document}

\ProvideTextCommandDefault{\textonehalf}{${}^1\!/\!{}_2\ $}

\title{Charge Trapping Correction and Energy Performance of the \MJD}
\newcommand{\ITEP}{National Research Center ``Kurchatov Institute'' Institute for Theoretical and Experimental Physics, Moscow, 117218 Russia}
\newcommand{\JINR}{Joint Institute for Nuclear Research, Dubna, 141980 Russia} 
\newcommand{\lbnl}{Nuclear Science Division, Lawrence Berkeley National Laboratory, Berkeley, CA 94720, USA}
\newcommand{\lbnle}{Engineering Division, Lawrence Berkeley National Laboratory, Berkeley, CA 94720, USA}
\newcommand{\lanl}{Los Alamos National Laboratory, Los Alamos, NM 87545, USA}
\newcommand{\queens}{Department of Physics, Engineering Physics and Astronomy, Queen's University, Kingston, ON K7L 3N6, Canada}
\newcommand{\uw}{Center for Experimental Nuclear Physics and Astrophysics, and Department of Physics, University of Washington, Seattle, WA 98195, USA}
\newcommand{\unc}{Department of Physics and Astronomy, University of North Carolina, Chapel Hill, NC 27514, USA}
\newcommand{\duke}{Department of Physics, Duke University, Durham, NC 27708, USA}
\newcommand{\ncsu}{Department of Physics, North Carolina State University, Raleigh, NC 27695, USA}	
\newcommand{\ornl}{Oak Ridge National Laboratory, Oak Ridge, TN 37830, USA}
\newcommand{\ou}{Research Center for Nuclear Physics, Osaka University, Ibaraki, Osaka 567-0047, Japan}
\newcommand{\pnnl}{Pacific Northwest National Laboratory, Richland, WA 99354, USA}
\newcommand{\ttu}{Tennessee Tech University, Cookeville, TN 38505, USA}
\newcommand{\sdsmt}{South Dakota Mines, Rapid City, SD 57701, USA}
\newcommand{\usc}{Department of Physics and Astronomy, University of South Carolina, Columbia, SC 29208, USA}
\newcommand{\usd}{Department of Physics, University of South Dakota, Vermillion, SD 57069, USA}  
\newcommand{\ut}{Department of Physics and Astronomy, University of Tennessee, Knoxville, TN 37916, USA}
\newcommand{\tunl}{Triangle Universities Nuclear Laboratory, Durham, NC 27708, USA}
\newcommand{\mpi}{Max-Planck-Institut f\"{u}r Physik, M\"{u}nchen, 80805, Germany}
\newcommand{\tum}{Physik Department and Excellence Cluster Universe, Technische Universit\"{a}t, M\"{u}nchen, 85748 Germany}
\newcommand{\williams}{Physics Department, Williams College, Williamstown, MA 01267, USA}
\newcommand{\ciemat}{Centro de Investigaciones Energ\'{e}ticas, Medioambientales y Tecnol\'{o}gicas, CIEMAT 28040, Madrid, Spain}
\newcommand{\iu}{Department of Physics, Indiana University, Bloomington, IN 47405, USA}
\newcommand{\iuceem}{IU Center for Exploration of Energy and Matter, Bloomington, IN 47408, USA}

\author{I.J.~Arnquist}\affiliation{\pnnl} 
\author{F.T.~Avignone~III}\affiliation{\usc}\affiliation{\ornl}
\author{A.S.~Barabash}\affiliation{\ITEP}
\author{C.J.~Barton}\affiliation{\usd}	
\author{K.H.~Bhimani}\affiliation{\unc}\affiliation{\tunl} 
\author{E.~Blalock}\affiliation{\ncsu}\affiliation{\tunl} 
\author{B.~Bos}\affiliation{\unc}\affiliation{\tunl} 
\author{M.~Busch}\affiliation{\duke}\affiliation{\tunl}	
\author{M.~Buuck}\altaffiliation{Present address: SLAC National Accelerator Laboratory, Menlo Park, CA 94025, USA}\affiliation{\uw} 
\author{T.S.~Caldwell}\affiliation{\unc}\affiliation{\tunl}	
\author{Y-D.~Chan}\affiliation{\lbnl}
\author{C.D.~Christofferson}\affiliation{\sdsmt} 
\author{P.-H.~Chu}\affiliation{\lanl} 
\author{M.L.~Clark}\affiliation{\unc}\affiliation{\tunl} 
\author{C.~Cuesta}\affiliation{\ciemat}	
\author{J.A.~Detwiler}\affiliation{\uw}	
\author{Yu.~Efremenko}\affiliation{\ut}\affiliation{\ornl}
\author{H.~Ejiri}\affiliation{\ou}
\author{S.R.~Elliott}\affiliation{\lanl}
\author{G.K.~Giovanetti}\affiliation{\williams}  
\author{M.P.~Green}\affiliation{\ncsu}\affiliation{\tunl}\affiliation{\ornl}   
\author{J.~Gruszko}\affiliation{\unc}\affiliation{\tunl} 
\author{I.S.~Guinn}\affiliation{\unc}\affiliation{\tunl} 
\author{V.E.~Guiseppe}\affiliation{\ornl}	
\author{C.R.~Haufe}\affiliation{\unc}\affiliation{\tunl}	
\author{R.~Henning}\affiliation{\unc}\affiliation{\tunl}
\author{D.~Hervas~Aguilar}\affiliation{\unc}\affiliation{\tunl} 
\author{E.W.~Hoppe}\affiliation{\pnnl}
\author{A.~Hostiuc}\affiliation{\uw} 
\author{M.F.~Kidd}\affiliation{\ttu}	
\author{I.~Kim}\affiliation{\lanl} 
\author{R.T.~Kouzes}\affiliation{\pnnl}
\author{T.E.~Lannen~V}\affiliation{\usc} 
\author{A.~Li}\affiliation{\unc}\affiliation{\tunl} 
\author{J.M. L\'opez-Casta\~no}\affiliation{\ornl} 
\author{E.L.~Martin}\altaffiliation{Present address: Duke University, Durham, NC 27708}\affiliation{\unc}\affiliation{\tunl}	
\author{R.D.~Martin}\affiliation{\queens}	
\author{R.~Massarczyk}\affiliation{\lanl}		
\author{S.J.~Meijer}\affiliation{\lanl}	
\author{S.~Mertens}\affiliation{\mpi}\affiliation{\tum}		
\author{T.K.~Oli}\affiliation{\usd}  
\author{G.~Othman}\altaffiliation{Present address: Universit{\"a}t Hamburg, Institut f{\"u}r Experimentalphysik, Hamburg, Germany}\affiliation{\unc}\affiliation{\tunl} 
\author{L.S.~Paudel}\affiliation{\usd} 
\author{W.~Pettus}\affiliation{\iu}\affiliation{\iuceem}	
\author{A.W.P.~Poon}\affiliation{\lbnl}
\author{D.C.~Radford}\affiliation{\ornl}
\author{A.L.~Reine}\affiliation{\unc}\affiliation{\tunl}	
\author{K.~Rielage}\affiliation{\lanl}
\author{N.W.~Ruof}\affiliation{\uw}	
\author{D.C.~Schaper}\affiliation{\lanl} 
\author{D.~Tedeschi}\affiliation{\usc}		
\author{R.L.~Varner}\affiliation{\ornl}  
\author{S.~Vasilyev}\affiliation{\JINR}	
\author{J.F.~Wilkerson}\affiliation{\unc}\affiliation{\tunl}\affiliation{\ornl}    
\author{C.~Wiseman}\affiliation{\uw}		
\author{W.~Xu}\affiliation{\usd} 
\author{C.-H.~Yu}\affiliation{\ornl}
\author{B.X.~Zhu}\altaffiliation{Present address: Jet Propulsion Laboratory, California Institute of Technology, Pasadena, CA 91109, USA}\affiliation{\lanl} 

\collaboration{{\sc{Majorana}} Collaboration}
\noaffiliation
\date{\today}

\begin{abstract}
$P$-type point contact (PPC) high-purity germanium detectors are an important technology in astroparticle and nuclear physics due to their superb energy resolution, low noise, and pulse shape discrimination capabilities.
Analysis of data from the \MJD, a neutrinoless double-$beta$ decay experiment deploying PPC detectors enriched in \nuc{76}{Ge}, has led to several novel improvements in the analysis of PPC signals.
In this work we discuss charge trapping in PPC detectors and its effect on energy resolution.
Small dislocations or impurities in the crystal lattice result in trapping of charge carriers from an ionization event of interest, attenuating the signal and degrading the measured energy.
We present a modified digital pole-zero correction to the signal energy
estimation that counters the effects of charge trapping and improves the energy
resolution of the \MJD~by approximately 30$\%$ to around 2.4 keV full width
at half-maximum at 2039 keV, the $^{76}$Ge $Q$ value. An alternative approach achieving similar resolution enhancement 
is also presented. 

\end{abstract}
\maketitle

\section{Introduction}
\label{sec:intro}
    Germanium semiconductor detectors' excellent energy resolution has made them the technology of choice for radiation detection and spectroscopy for decades. High resolution is particularly advantageous in searches for neutrinoless double-beta decay (\nonubb)~\cite{Agostini:2022zub,Avignone:2019phg,Dolinski2019,Barabash2018,Vergados2016,DelloOro2016}, a postulated process beyond the standard model in which two neutrons in a nucleus transform into two protons and emit two electrons but no antineutrinos, violating lepton number conservation. 
    Experimental observation of \nonubb\ decay would provide essential insights into the observed matter-antimatter asymmetry of the universe. The existence of this process would imply that neutrinos are Majorana particles, i.e., their own antiparticles~\cite{Schechter:1982}.
    Current-generation \nonubb\ decay experiments have recently placed lower limits on the \nonubb\ half-life in various isotopes that in some cases extend beyond $10^{26}$ years~\cite{Azzolini:2019tta,Agostini:2020xta,Arnquist:2022zrp,Armengaud:2020luj,CUORE:2021mvw,KamLAND-Zen:2022tow,Anton:2019wmi}, while some next-generation experiments aim to reach sensitivities of $10^{27-28}$ years to test the inverted neutrino mass ordering~\cite{Agostini:2021kba,LEGEND:2021bnm, nEXO:2021ujk,CUPIDInterestGroup:2019inu}. To maximize the discovery potential, these experiments require extremely low levels of radioactive background, a large active detector mass, and sufficient energy resolution. The latter is required both to distinguish the \nonubb\ peak from the irreducible \twonubb\ background, as well as to minimize the impact of continuum backgrounds.

\begin{figure}[!htbp]
        \includegraphics[width=\columnwidth]{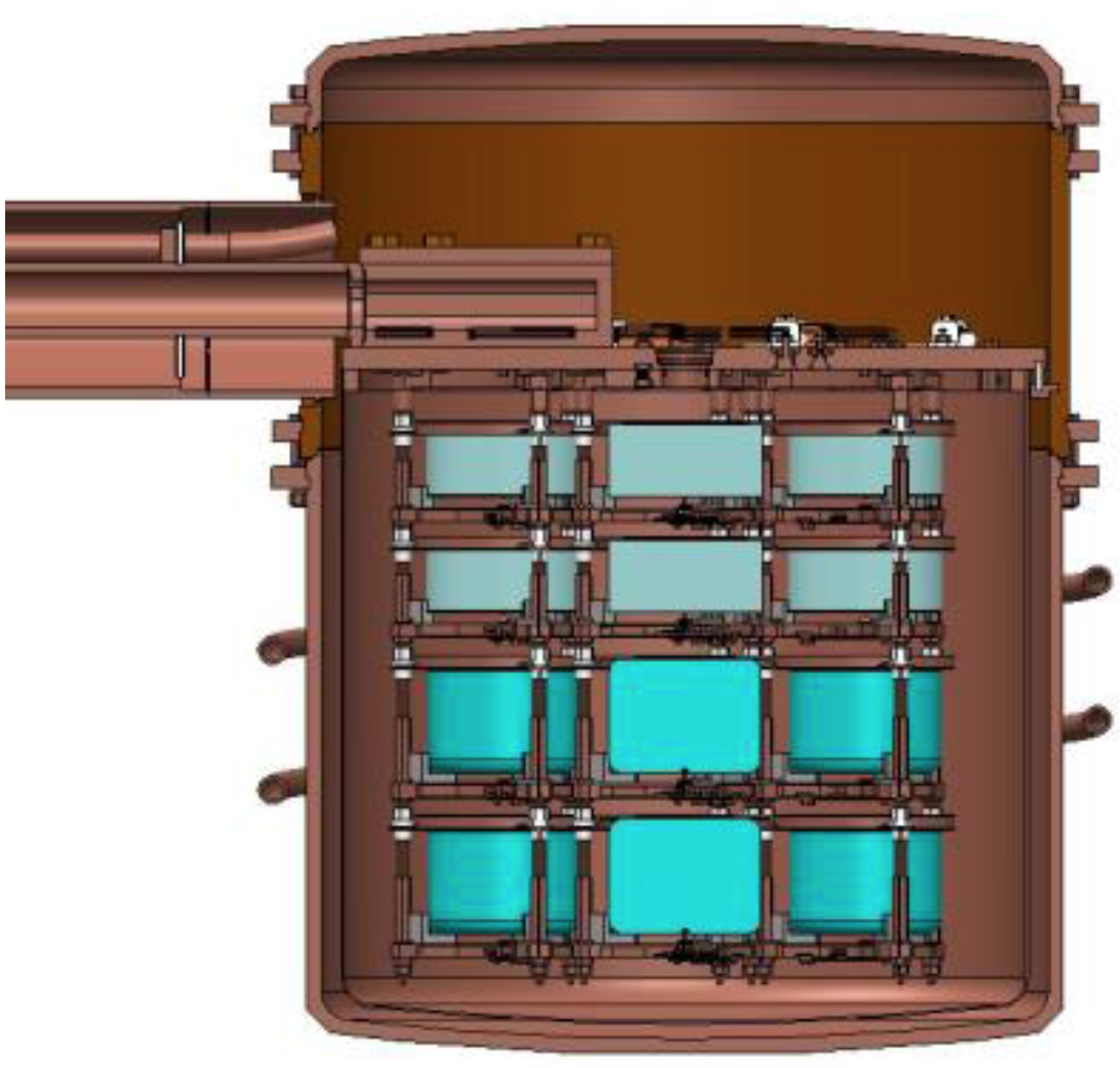}
        \caption{Rendering of one of the two \MJD\ vacuum cryostats containing a close-packed array of AMETEK-ORTEC PPC (cyan) as well as Canberra BEGe (grey blue) HPGe detectors. The primary structural material is ultra-pure copper electroformed underground at SURF. A thermosyphon provides cooling to liquid nitrogen temperatures via the cross-arm extending to the left, which is also the path for signal and HV cabling to penetrate the copper, lead, and plastic shielding that surrounds the cryostats. The height of the cryostat is about 46 cm.}
        \label{fig:modules}
    \end{figure}

    The \MJD~\cite{Abgrall:2014} is aimed at searching for \nonubb\ decay using high-purity germanium (HPGe) detectors enriched to 88$\%$ \nuc{76}{Ge}~\cite{ABGRALL2018314}. The experiment is sited at the 4850-foot level of the Sanford Underground Research Facility (SURF)~\cite{Heise_2015} in Lead, SD. Initially 29.7~kg of enriched Ge was deployed along with 14.4 kg of natural-abundance Ge detectors, with 58 individual detector units deployed in two vacuum cryostats. All enriched Ge detectors are $P$-type point contact (PPC) HPGe detectors manufactured by AMETEK-ORTEC Inc.~\cite{ortec} while all natural Ge detectors are broad energy germanium (BEGe) detectors manufactured by Canberra Industries~\cite{canberrabege}. 
    Each cryostat contains seven detector strings, illustrated in Fig.~\ref{fig:modules}. In 2020 the apparatus was upgraded, with 5.41~kg of enriched PPC HPGe detectors replaced with 6.67~kg of inverted-coax point-contact (ICPC) detectors manufactured by AMETEK-ORTEC Inc.~\cite{Cooper2011} The ICPC technology is being pursued as part of the research and development program for the LEGEND experiment. This paper will focus on data from the \MJD's PPC detectors.

    The \MJD\ achieved the best energy resolution of all large-scale \nonubb\ decay experiments to date~\cite{Aalseth:2017,Alvis:2019sil,Arnquist:2022zrp}. A critical step in this achievement was the implementation of a correction for energy degradation due to charge trapping. 
    Energy is measured in HPGe detectors by collecting and measuring the charge carriers liberated in ionization events. As the charge carriers drift in the detectors' internal fields, they
    produce signals sensitive to charge trapping  effects. Charge carriers can be trapped if they encounter impurities or dislocations in the crystal lattice. If they are not released on a timescale that is short compared to the signal collection time, the signal amplitude, and hence the energy, will be degraded. When the energy degradation is on the order of typical fluctuations in charge carrier generation, as it is in the \MJD, its net effect in a population of signals with varying collection times is to widen the detector response function.  Correcting for this trapping can thus result in improved energy reconstruction, as has been demonstrated for example in planar germanium detectors~\cite{HULL2014125}.

    In this paper, we discuss novel digital signal processing algorithms used to correct for charge trapping during reconstruction of event energy in PPC detectors. We then evaluate the algorithms' performance using data collected with detectors operating in the full \MJD\ array and in the ``string test cryostats" used for detector characterization and acceptance tests. In the string test cryostats, detector temperatures are estimated to be $\approx$95~K, roughly 15~K above the operating temperature of the full array. The strong temperature dependence in release times resulted in different charge trapping effects in these two systems. 
    
\section{Charge Trapping in PPC Detectors and Effective Pole-Zero Correction}
\label{sec:trapping}
    HPGe detectors have a ``p-i-n'' structure, where a large crystal of nearly intrinsic p- or n-type material is modified by the addition of p$^+$ and n$^+$ contacts on its surface. In the \MJD\ PPC detectors, the p-type impurity concentration in the bulk material is typically on the order of $10^{10}$ atoms/cm$^3$. The p$^+$ contact is created by boron implantation in a small ``point contact'' on one face of the crystal. The n$^+$ contact is created by lithium diffusion and covers the opposite face and cylindrical side of the crystal. A passivated surface insulates the boundary between contacts, and a 1--5 kV reverse bias voltage between the contacts depletes the intrinsic region of any free charge carriers.
    Any ionizing particle or process within the depletion region will create charge clouds of electron-hole pairs that are respectively transported to the n and p sides of the junction by the combined electric field of the bias voltage and the charge impurity gradient. The electrical signal produced at the n$^+$ and p$^+$ contacts can be understood in terms of the Shockley-Ramo theorem~\cite{shockley1938, Ramo:1939vr}, which states that the 
    amplitude of the charge signal induced by a moving carrier depends on the change in the weighting potential between its initial and final positions. The weighting potential depends only on the geometry of the electrode configuration and is obtained by setting the contact of interest to unit voltage and grounding all other contacts. 
    
\begin{figure}[!htbp]
        \includegraphics[width=\columnwidth]{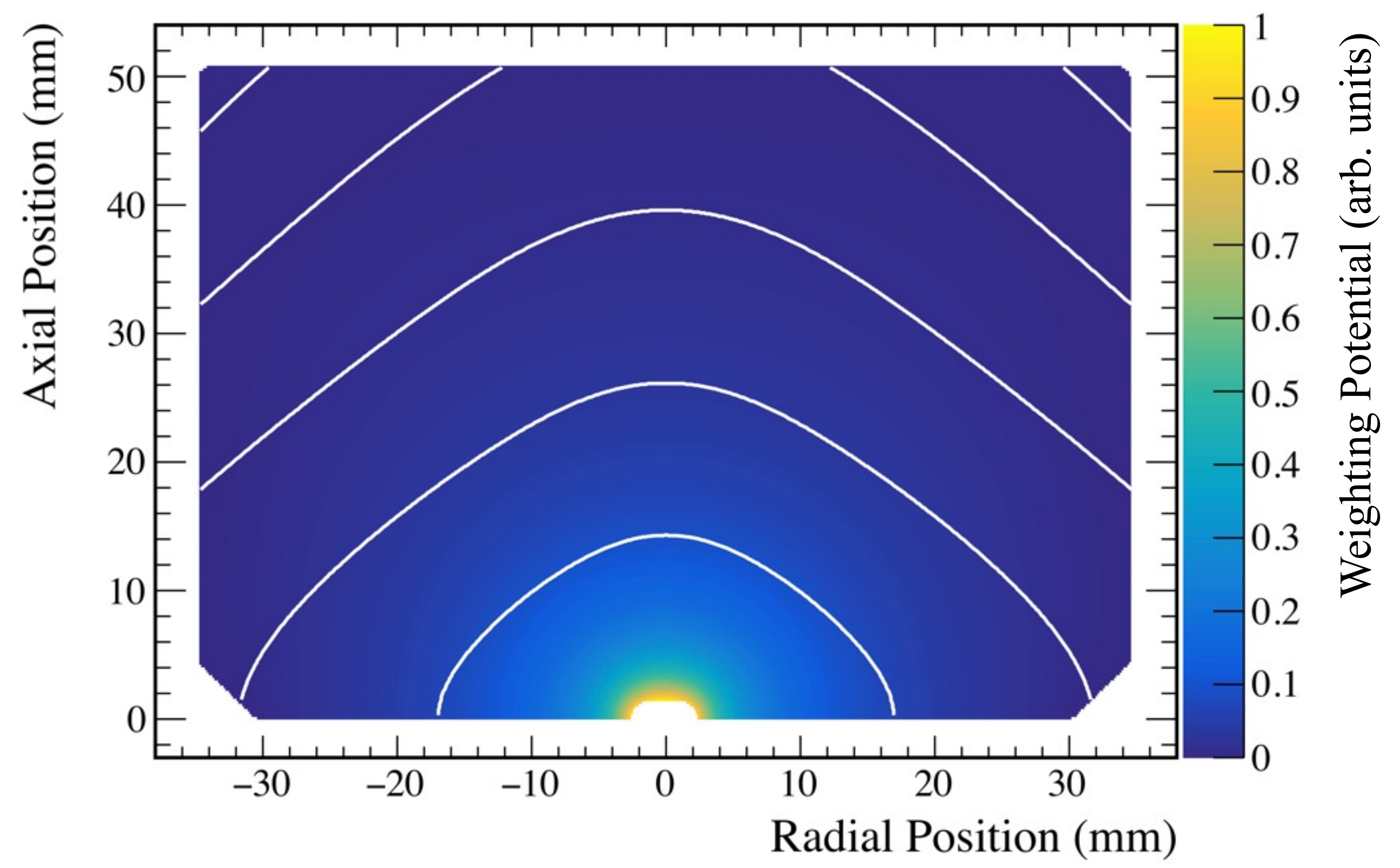}
        \caption{Simulated weighting potential in a PPC detector, showing its localization near the point contact region at the bottom center. The lithium diffused surface covers the top and the side region of the detector. The passivated surface surrounds the point contact at the bottom region. The white lines are isochrones of equal drift time for holes to reach the point contact spaced by 200 ns. Figure from~Ref.~\cite{Alvis:2019sil}.}
        \label{fig:weighting}
    \end{figure}

    PPC detectors have a sharply peaked weighting potential near the point contact, as shown in Fig.~\ref{fig:weighting}. Most electron-hole pairs are generated at a distance from the point contact. Since only the holes traverse the region with high weighting potential, PPC signals tend to be hole-dominated. The shape and timing of the induced signal depends on charge carrier drift paths, so a great deal of information can be extracted from each pulse~\cite{Alvis:2019sil}. The entire signal formation process can be simulated to high numerical accuracy, allowing detailed comparisons of simulation and data to explore subtle effects of detector properties on pulse shape. In this work, we use a version of the HPGe pulse shape simulation package \texttt{siggen} specialized for \MJD\ detectors~\cite{Radford:2014}.

\begin{figure}[!htbp]
        \centering
        \includegraphics[width=\columnwidth]{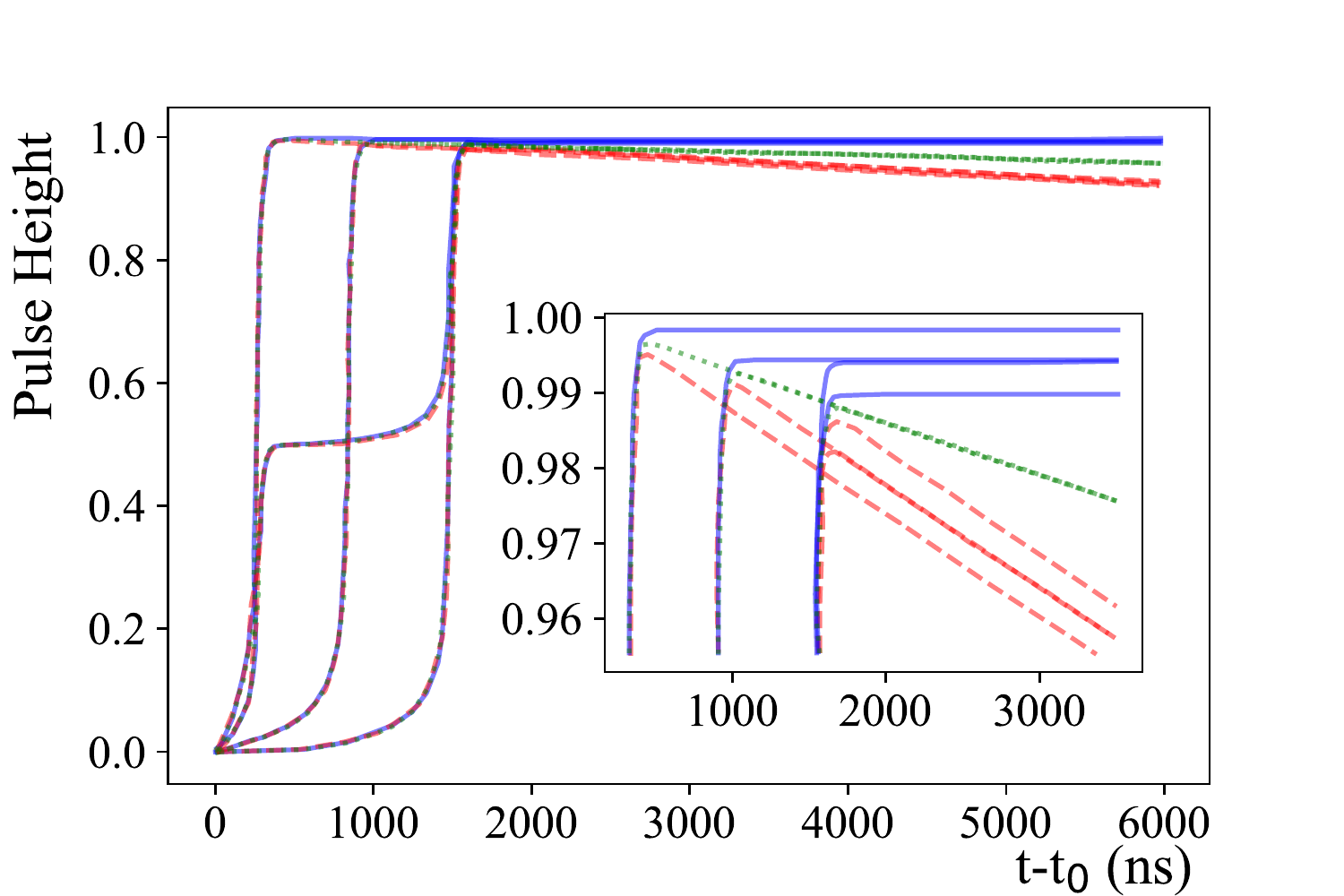}
        \caption{Simulated PPC waveforms from \texttt{siggen} with exaggerated drift times and preamplifier decay. Red dashed: no correction (raw signal), blue solid: standard $\tau_{RC}$ correction, green dotted: effective $\tau$ correction. The pulse with a large step in the rising edge is due to a ``multisite'' event in which simultaneous energy depositions occur at multiple positions in the detector. The inset zooms in on the upper left region of the figure.}
        \label{fig:ctsim}
    \end{figure}

    Charge carriers in events undergoing longer drift have an increased probability to encounter traps from impurities or dislocations in the crystal lattice. The trapped carriers are subsequently released at a rate dependent on the effective depth of the trap and the detector temperature. The release time increases with trap depth. 
    As the temperature of the detector is lowered, the release time can become long compared to the signal collection time, leading to pulses with degraded amplitude.
    In this regime, under the assumption that the trapping probability is independent of location within the crystal, a constant Poisson probability for charge trapping per unit time during the charge drift leads to exponential charge loss, with time constant $\tau_{\rm ct}$. As charges are trapped, their drift ceases to contribute to the signal amplitude. This amplitude degradation is visualized in the blue simulated single-site pulses in Fig.~\ref{fig:ctsim}, which are generated with unit amplitude at detector positions with short, intermediate, and long drift times. Although we did not rigorously validate our assumption of trapping probability uniformity, this exponential decay model yielded excellent results here.
    
    In the \MJD, detector signals are read out using charge sensitive preamplifiers with a resistive feedback loop~\cite{Majorana:2021mtz}. Pulse traces are recorded with a 14-bit 100~MHz digitizer, capturing 20--50~$\mu$s of data. 
    AC coupling in the second-stage amplification gives the recorded signals an exponential decay constant of $\tau_{RC} \approx 72~\mu$s, as simulated in the red traces in Fig.~\ref{fig:ctsim}. A standard pole-zero correction~\cite{Leo:1994}, applied offline as a digital filter, can correct for this decay in the digitized signals prior to trapezoidal filtering to estimate the pulse amplitude~\cite{Jordanov:1994}. However, such fully pole-zero corrected traces simply recover the situation shown in the blue traces in Fig.~\ref{fig:ctsim}, exhibiting significant amplitude dependence on drift time. As a result, detector energy resolution can be heavily degraded.
    
    If drift times can be extracted directly, energies can be manually corrected to reduce this detector resolution degradation~\cite{Kephart:2009, Martin:2012}. In practice, drift time can be difficult to extract, and algorithms can be sensitive to various kinds of noise or other anomalies or exhibit energy- as well as drift time-dependent biases that are hard to avoid. In addition, because the drifting charges contribute to the signal the entire time they are drifting, the degradation is dependent not just on the drift time, but also on the trajectories of the drifting charges.
    
    In this work we implemented a different energy reconstruction strategy that corrects for the signal degradation more directly. Instead of fully pole-zero correcting the pulses, an ``effective'' pole-zero correction is performed using a time constant ($\tau$) given by the difference between the charge-trapping and resistive-capacitive (RC) decay constants:
    \begin{equation}
        \frac{1}{\tau} \equiv \frac{1}{\tau_{RC}} - \frac{1}{\tau_{\rm ct}}.
        \label{eqn:taueff}
    \end{equation}
    As described in detail in Appendix~\ref{app:signal}, this correction effectively deconvolves the electronics response from the waveform, and then convolves it with an exponential response governed by $\tau_{\rm ct}$. The latter convolution has the effect of adding back in, sample by sample in the waveform, the charge lost due to trapping during each digitization time step.

    Post-filtering, the pulses exhibit exponentially decaying tails with time constants equivalent to $\tau_{\rm ct}$, as shown in the green traces in Fig.~\ref{fig:ctsim}.
    The alignment of the tails of these traces demonstrates that for any fixed time interval following the onset of the pulse, all pulses lose the same fractional amplitude. We thus first extract the pulse onset time ($t_0$), and then evaluate the energy of the pulse from the amplitude of the trapezoidal-filtered effective-pole-zero-corrected trace at a fixed time interval following $t_0$.


\section{Energy and $t_0$ Evaluation}
\label{sec:energy}

    The energy estimation of \MJD~events is done in two stages. An uncalibrated energy is determined by applying a recursive trapezoidal filter~\cite{Jordanov:1994}. The uncalibrated energy is then converted to keV based on fits to peaks of known energy in calibration spectra. A trapezoidal filter includes a ramp, a flat-top, and a falling region with durations $t_{\text{ramp}}$, $t_{\text{flat}}$, and $t_{\text{fall}}$, respectively.
    The {\it ramp} and {\it fall} times are selected to give optimal estimates of the voltage levels before and after the incident pulse, and the {\it flat} time is chosen to be slightly longer than the longest drift times in the detector.
    For symmetric trapezoidal filters, $t_{\text{fall}} = t_{\text{ramp}}$. 
    We will denote the rise, flat-top, and fall times of trapezoidal filters using the notation [{\it rise, flat, fall}\,], where each time is given in units of $\mu$s. 
    
\begin{figure}[!htbp]
        \includegraphics[width=\linewidth]{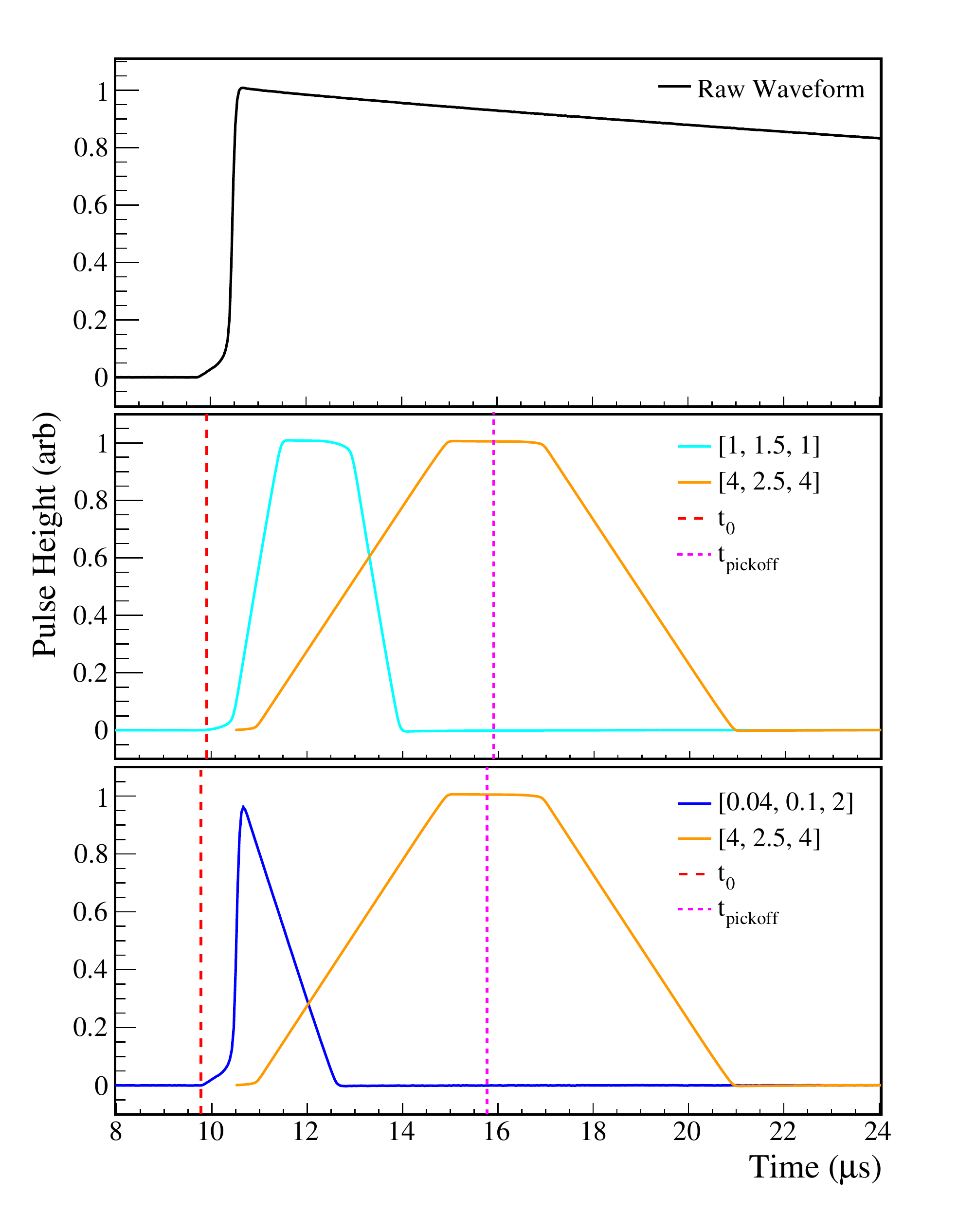}
        \caption{The fixed-time-pickoff technique used to estimate the uncalibrated energy.  Top: A normalized raw waveform from data.  Middle: Symmetric leading-edge (cyan) and energy (orange) trapezoidal-filtered signals. Bottom: Asymmetric leading-edge (blue) and energy (orange) trapezoidal-filtered signals. The filters are labeled using the convention $[rise, flat, fall]$ in $\mu$s.  The dashed lines show $t_0$, and the dotted lines indicate the energy pickoff time. Note that the rising edge of the asymmetric filter more closely resembles the rising edge of the waveform, producing a better $t_0$ estimation.}
        \label{fig:ftp}
    \end{figure}

    In contrast with the common method of using the maximum value of a symmetric trapezoidal filter as the uncalibrated energy estimate, the \MJD~ uses a ``fixed-time pickoff'' technique in which we first determine the start time of the pulse in the filtered signal, $t_0$, and then the uncalibrated energy is taken to be the value of the filtered signal at the fixed time $t_0 + \delta t$, as shown in Fig.~\ref{fig:ftp}. 
    The advantage of the fixed-time pickoff is two-fold. First, the noise in this energy estimate is governed by the statistics of the single picked-off sample, rather than the maximum of a series of samples, and thus avoids the bias to higher energy of the maximum value estimator. This bias is relevant for low-energy signals (below $\approx$10~keV).
    Second, careful selection of the short-rise-time trapezoidal filter's parameters can provide an accurate $t_0$ time that naturally encodes the drift time of the signal. By combining this fixed-time pickoff with the effective pole-zero correction described in Sec.~\ref{sec:trapping}, an optimized energy estimator is obtained with a free parameter $\tau$ that can be tuned for individual detectors.
    

    The start time $t_0$ of each pulse is evaluated using two different methods, each using a leading-edge algorithm applied to a filtered signal.
    Our initial $t_0$ algorithm, used in Refs.~\cite{Aalseth:2017} and~\cite{Alvis:2019sil}, used a trapezoidal filter with a symmetric shape of $[1, 1.5, 1]$.
    The second algorithm, used in Ref.~\cite{Arnquist:2022zrp}, uses an asymmetric trapezoidal filter with a shape of $[0.04, 0.1, 2]$. For both filters, the pulse start time
    $t_0$ is identified by stepping backwards from the maximum of the filtered signal until the first crossing of a very low (2~ADC~$\approx1$~keV) threshold, and then interpolating between the samples before and after the threshold crossing.
    Since they rely on a threshold crossing, both algorithms result in a bias toward later times (discussed in Sec.~\ref{sec:t0error}). However the asymmetric $[0.04, 0.1, 2]$ trapezoidal filter results in a more accurate and precise $t_0$ determination, since it is more optimized to preserve timing information in the rising edge, while still minimizing the impact of both low and high frequency noise.
    Figure~\ref{fig:ftp} shows the symmetric and asymmetric filters, as well as the $t_0$ values determined by these algorithms.
    
   After obtaining $t_0$, we then use a $[4, 2.5, 4]$ trapezoidal filter to evaluate the uncalibrated energy with a fixed-time pickoff at $t_0 + \delta t$, where
    \begin{align}
        \label{eqn:offset}
        \begin{split}
        \delta t =\ &  (t_{\mathrm{ramp}} + t_{\mathrm{flat}}) - 0.5~\mu\text{s}.
        \end{split}
    \end{align}
   In this equation, the time constants refer to those of the $[4, 2.5, 4]$ energy filter. The value 0.5 $\mu$s was determined empirically to assure that the pickoff time is in the flat top of the filtered signal before the falling edge as shown in Fig.~\ref{fig:ftp}. 


\section{Energy Resolution Optimization}
\label{sec:resolution}
    The \MJD~data are divided into data sets (DS) separated by hardware or software-based changes to the experimental configuration and are described in Ref.~\cite{Alvis:2019sil}.
    The analysis in this paper primarily uses data taken during weekly  \nuc{228}{Th} source calibrations~\cite{mjdcalibration2017}.
    Waveform data for each physics event from the data acquisition system is processed and stored for offline analysis.
    
    To obtain the energy of an event, the \MJD~uses the  $[4, 2.5, 4]$ trapezoidal filter with the effective pole-zero correction and fixed-time pickoff energy estimator described earlier. 
    Calibration constants determined using \nuc{228}{Th} data are then applied to convert the uncalibrated energy to keV. The full width at half-maximum (FWHM) of the \nuc{208}{Tl} peak at 2615~keV is measured by fitting that peak with a peak shape function consisting of a Gaussian, a step function (an erf), a low-energy tail modeled by an exponentially modified Gaussian, and a quadratic background~\cite{Alvis:2019sil}.
    
\begin{figure}[!htbp]
        \centering
        \includegraphics[width=\columnwidth]{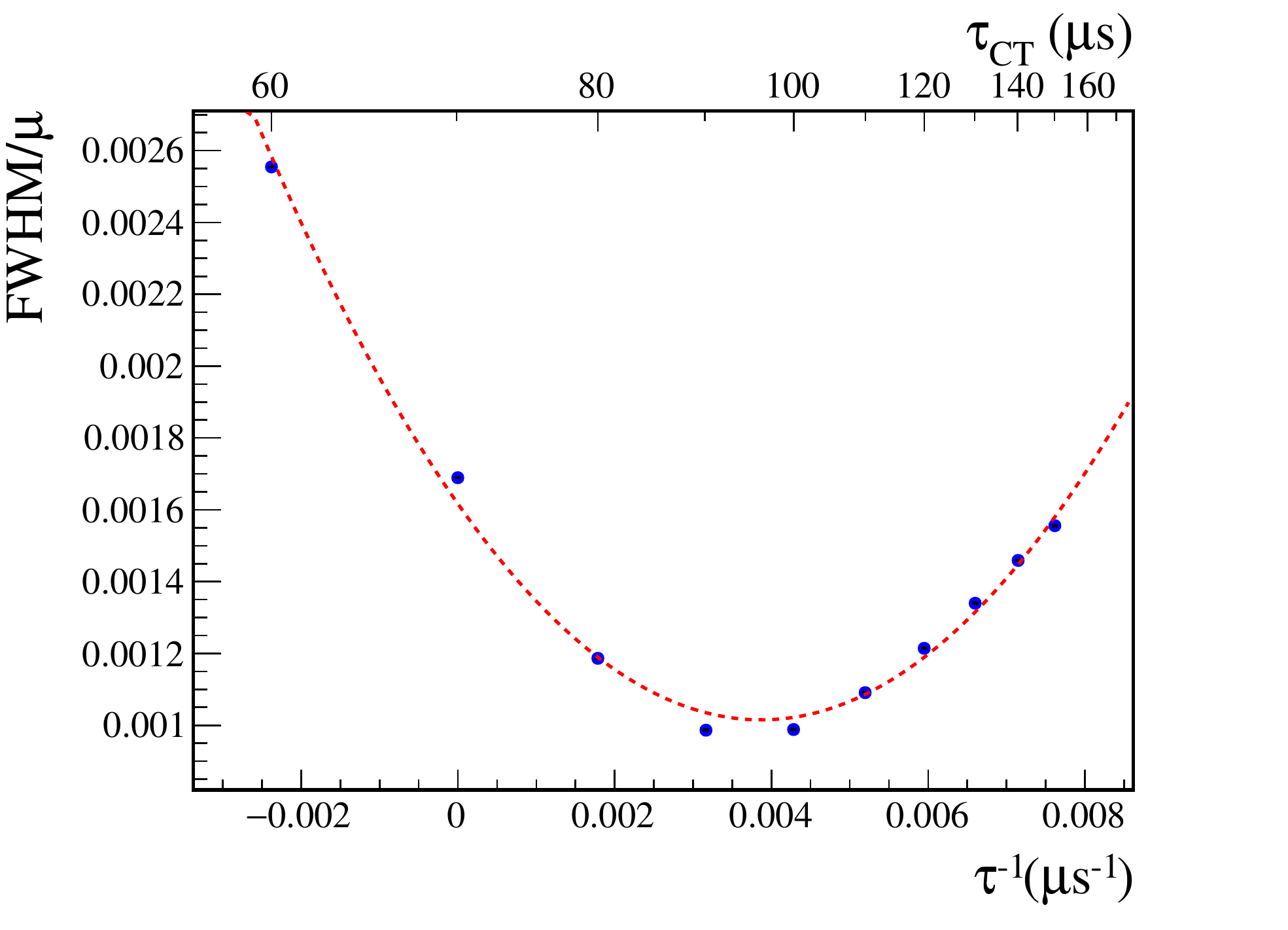}
        \caption{FWHM/$\mu$ versus $1/\tau$ fit to a quadratic function. For this detector, a clear minimum is found at $1/\tau \approx 0.004$ which is close to the point where $\tau_{\rm ct} = 100 \mu$s using Eq.~\ref{eqn:taueff}.}
        \label{fig:quad_112_DS0}
    \end{figure}

    In the \MJD~algorithm, the effective time constant ($\tau$) that optimizes the energy resolution of the 2615~keV peak is determined detector by detector for each data set. This is done by varying $\tau_{\rm ct}$ in 10~$\mu$s steps from 60 to 250~$\mu$s, determining the FWHM of the 2615 keV peak at each value of $\tau$, and then fitting FWHM/$\mu$ versus $\tau^{-1}$ to a quadratic function, where $\mu$ is the fit peak position:
    \begin{equation}
        \mathrm{FWHM}/\mu = p_0 + p_1\ \tau^{-1} + p_2\ \tau^{-2}.
    \end{equation}
    This optimization process is illustrated in Fig.~\ref{fig:quad_112_DS0} for detector P42575A using a 1-h \nuc{228}{Th} calibration. For this detector, a clear minimum is found near $\tau^{-1}\approx 0.004~\mu\mathrm{s}^{-1}$; we choose the numeric minimum corresponding to $\tau_{\rm ct} = 90 ~\mu$s according to Eq.~\ref{eqn:taueff}. This is a reasonable choice when the detector energy gain and operating bias voltage remain roughly constant over time, which is the case for most calibration periods in the \MJD, as shown below in Fig~\ref{fig:fwhmtime}. 

\begin{figure}[!htbp]
        \centering
        \includegraphics[width=0.95\columnwidth]{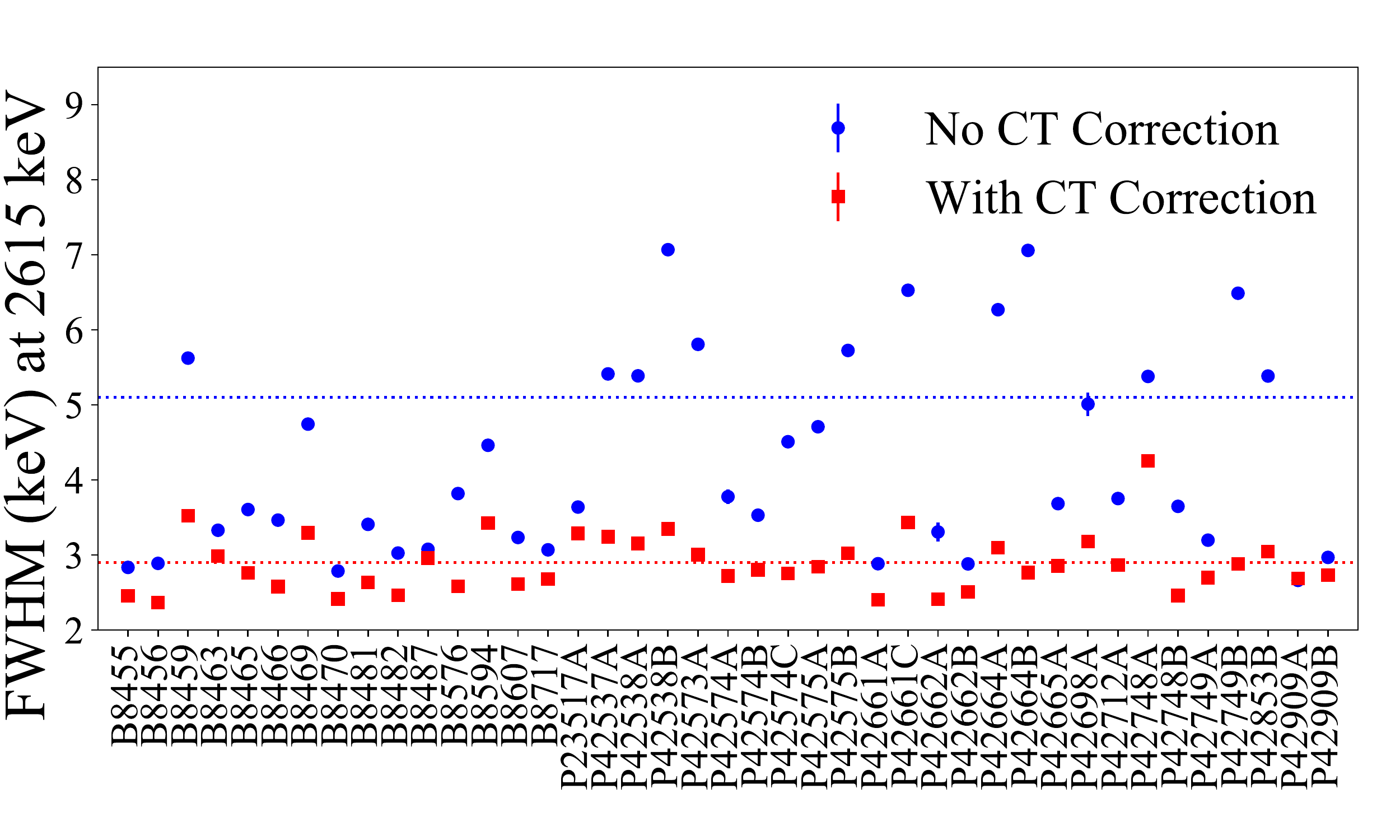}
        \caption{Energy resolution (FWHM, statistics only) at 2615~keV for all operating detectors in DS6a before (blue points) and after (red points) the charge trapping correction. The $x$-axis is the detector ID. Detector IDs beginning with ``B'' and ``P'' are natural and enriched Ge detectors, respectively. The average resolution after the charge trapping correction (red line) represents a $\approx$30\% improvement. The variation between detectors is mostly due to differing crystal impurities and maximum drift times. Figure from Ref.~\cite{Alvis:2019sil}. }
        \label{fig:fwhm_DS6a}
    \end{figure}

\begin{figure}[!htbp]
        \includegraphics[width=\columnwidth]{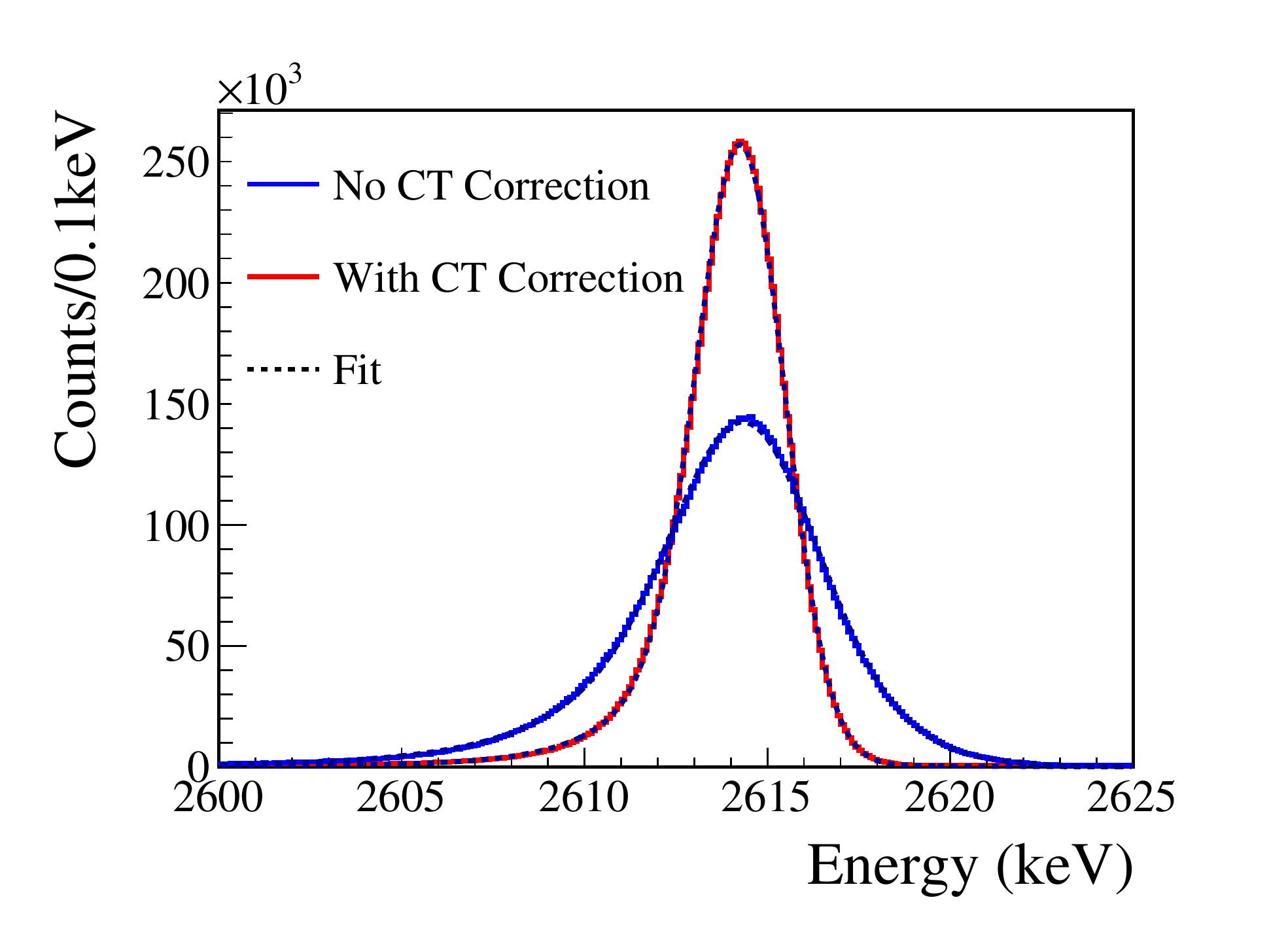}
        \caption{The 2615-keV peak from \nuc{208}{Tl} in calibration data with all operating enriched detectors in DS0--DS6a. The fits reproduce the measured peak shapes nearly perfectly.}
        \label{fig:2615peak}
    \end{figure}

\begin{figure}[!htbp]
        \includegraphics[width=\columnwidth]{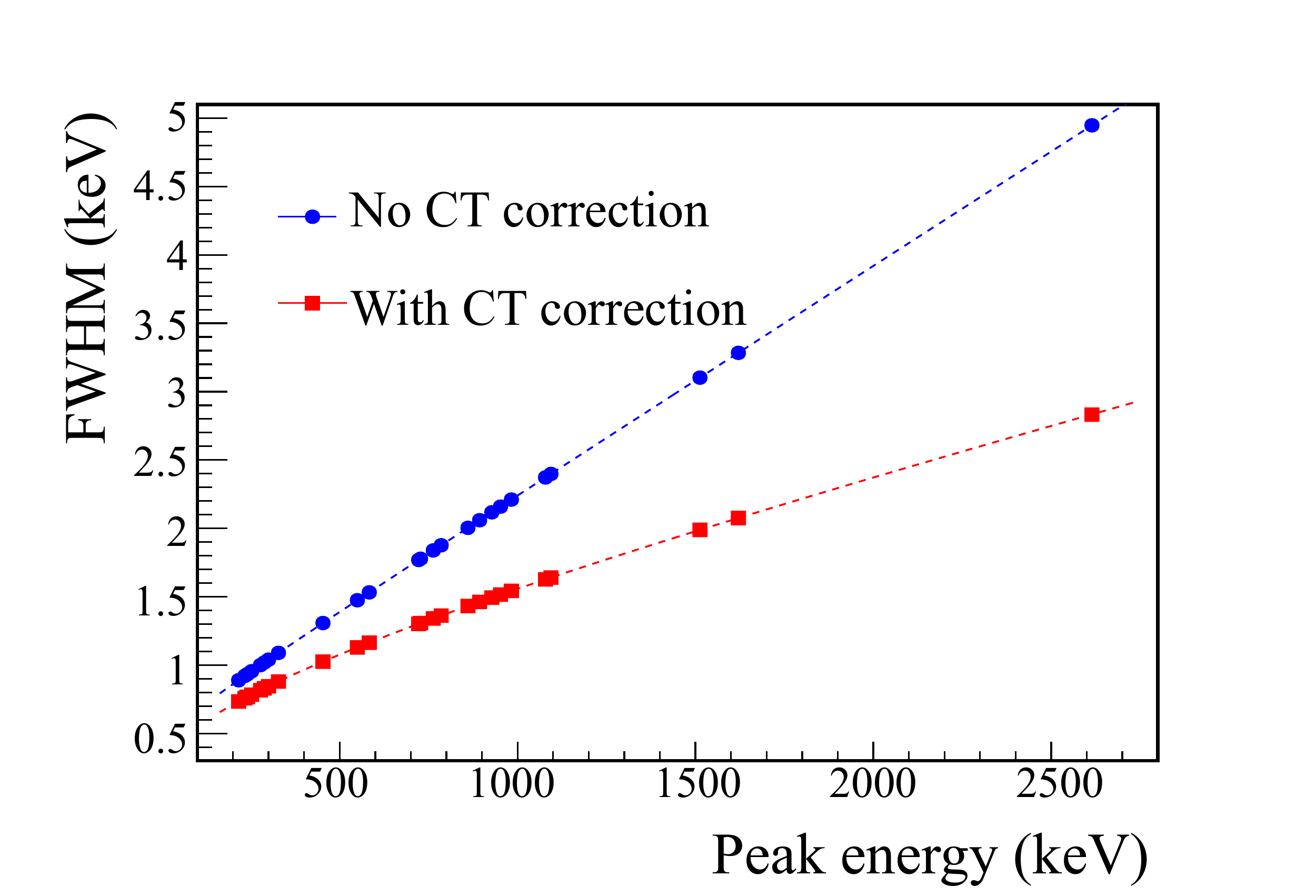}
        \caption{Resolution (FWHM, statistics only) vs energy for the combined energy spectrum of all operating detectors in DS0--DS6a with fits to Eq.~\ref{eq:sigfit}. The energy resolution of this curve at the $Q$ value is about 2.4 keV with the charge trapping correction. 
        Inclusion of systematic uncertainty increases the FWHM to 2.5~keV.}
        \label{fig:reso_DS05}
    \end{figure}

    This effective charge trapping correction results in an average 30$\%$ improvement in the FWHM at 2615 keV for all operating detectors in DS6a, as shown in Fig.~\ref{fig:fwhm_DS6a}. The energy resolution over the entire spectrum is parametrized by
    \begin{equation}
        \label{eq:sigfit}
        \sigma(E) = \sqrt{\sigma_n^2 + \sigma_f^2 E + \sigma_c^2 {E}^2}
    \end{equation}
    where $\sigma_n$ accounts for electronic noise, $\sigma_f$ accounts for the Fano factor~\cite{Fano:1947}, and $\sigma_c$ is a term resulting in a linear dependence at high energy that is dominated by residual charge trapping.
    The inclusion of all data incorporates statistical variation between calibrations, different detectors, and gain instabilities that can contribute to the $\sigma$ parameters; here we report the results using the statistics of the calibrations only, to highlight the impact of the charge trapping correction.
    Figure~\ref{fig:2615peak} shows the peak shape at 2615 keV for all operating enriched detectors combined. The charge trapping correction results in a markedly sharper peak, with FWHM improved from 5.1 keV to 2.9 keV.
    Figure~\ref{fig:reso_DS05} shows the energy resolution curve as a function of energy for the sum of all operating detectors in DS0--DS6a (the dataset reported in~\cite{Alvis:2019sil}) with and without the charge trapping correction applied. Prior to application of the charge trapping correction, resolution scales mostly linearly with energy due to the $\sigma_c$ term; with the charge trapping correction, the Fano term $\sigma_f$ dominates.
    Figure~\ref{fig:fwhmtime} shows the FWHM of the combined spectrum of each calibration subset as a function of time from DS0--DS6a. The instability in resolution seen during DS5 is due to noise introduced by imperfect grounding during the final stages of construction of the outer poly shield of the experiment. Finally, Table~\ref{tab:resolution_2039} lists the FWHM at the \nonubb~$Q$ value of 2039 keV determined by fits to Eq.~\ref{eq:sigfit}, for each data set published in Ref.~\cite{Alvis:2019sil}. There is no significant difference in the energy resolution of the enriched and natural detectors, which were manufactured by different companies with different designs, after the charge trapping correction. Both have an average resolution of about 2.2--2.4~keV FWHM without considering the uncertainties due to nonlinearities and gain stability, which increase the FWHM to 2.5~keV~\cite{Alvis:2019sil}. 

    \begin{figure}[!htpb]
        \includegraphics[width=\columnwidth]{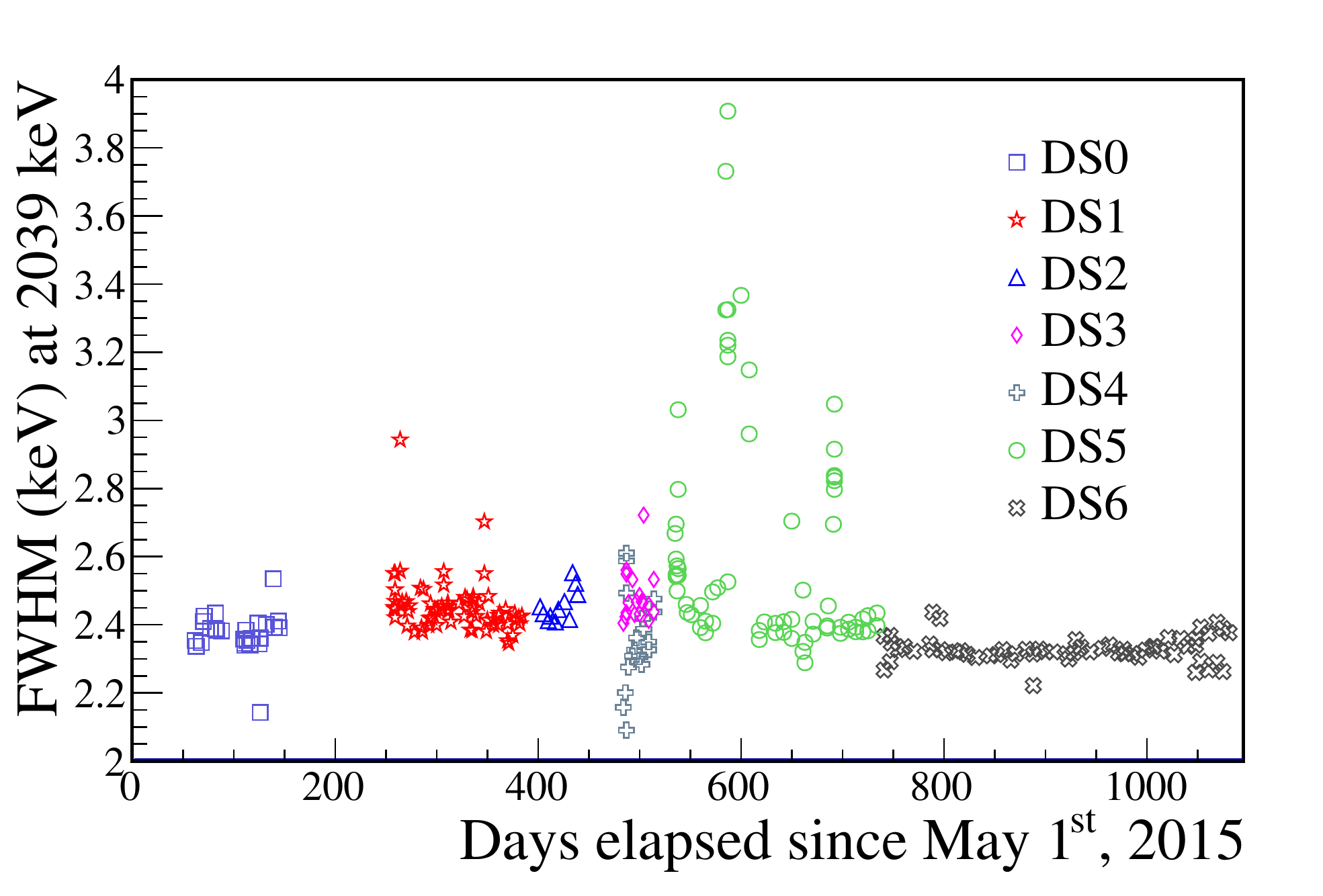}
        \caption{Energy resolution (FWHM, statistics only) of all operating detectors at 2039 keV versus time over each weekly calibrations in DS0--DS6a. The instability in DS5 is due to construction activities that induced higher noise.}
        \label{fig:fwhmtime}
    \end{figure}

    \begin{table}[!htbp]
        \caption{Optimized FWHM energy resolution (statistics only) in keV at 2039 keV for DS0--DS6a, showing enriched, natural, and all detectors combined, determined by fitting Eq.~\ref{eq:sigfit}. Systematic contributions raise the final combined FWHM to 2.5~keV.}
        \setlength{\tabcolsep}{12pt} 
        \begin{center}
            \begin{tabular}{c|c|c|c}
              \hline
              & $^{\mathrm{enr}}$Ge & $^{\mathrm{nat}}$Ge & All \\
              \hline \hline
              DS0 & 2.472(5) & 2.214(4)  & 2.403(5)  \\
              DS1 & 2.465(6) & 2.131(3)  &  2.441(5)  \\
              DS2 & 2.468(6) & 2.118(4)  & 2.444(6) \\
              DS3 & 2.477(6) & 2.295(4)  & 2.430(5) \\
              DS4 & 2.475(4) & 2.405(4)  & 2.456(4) \\
              DS5a & 2.815(7) & 2.508(4)& 2.727(6) \\
              DS5b & 2.434(5) & 2.256(4)  & 2.381(5) \\
              DS5c & 2.444(2) & 2.260(2) & 2.392(2) \\
              DS6a & 2.381(3) & 2.225(2) & 2.334(3) \\\hline
              DS0-DS6a & 2.484(5) & 2.299(3) & 2.438(5)\\
              \hline
          \end{tabular}
        \end{center}
        \label{tab:resolution_2039}
    \end{table}

\section{Temperature Dependence of Charge Trapping}
\label{sec:stc}

    Acceptance testing for \MJD\ PPC detectors was carried out underground at SURF both in vendor-provided cryostats as well as, for some detectors, in custom-built string test cryostats. The string test cryostats were designed to operate strings
    of 3--5 detectors using the same detector mounts, front ends, amplifiers, and readout electronics as the \MJD. 
    Charge trapping corrections were not applied for the acceptance tests, which still showed uniformly good energy resolution usually below 3~keV FWHM at 2615~keV and within $\approx$10\% of the values measured {\it in-situ} in the \MJD\ modules with optimized charge trapping corrections. 
    
\begin{figure}[!hbtp]
        \includegraphics[width=\columnwidth]{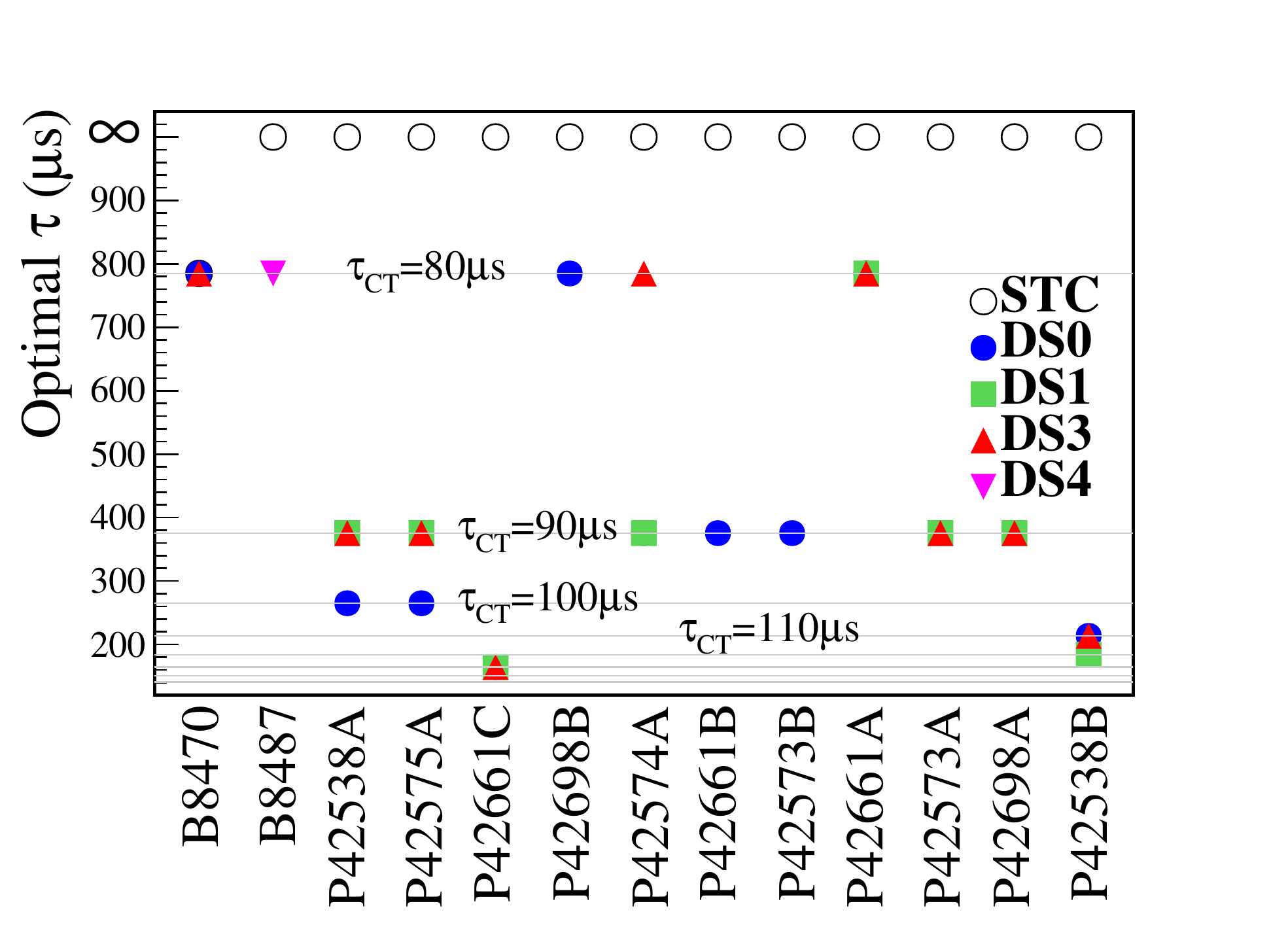}
        \caption{Optimal values of $\tau$ for data taken in string test cryostats (STC) and \MJD\ modules for 13 detectors. The $x$-axis is the detector ID; those starting with ``B'' are BEGe detectors, the remaining are enriched PPC detectors. Some symbols (such as the STC value for B8470 at $\tau = 800~\mu$s are not visible due to overlaps. Horizontal grey lines indicate the inferred values of $\tau_{\rm{ct}}$. The symbol $\infty$ indicates that charge trapping corrections do not improve the energy resolution, implying that there is little to no charge trapping. We find slight variations between data sets.}
        \label{fig:tdep}
    \end{figure}

    Thirteen detectors were operated in both string test cryostats and the \MJD\ modules, including two BEGe detectors fabricated from natural Ge. The string test cryostat data were reanalyzed including the charge trapping correction described above. The optimal values of $\tau_{\rm ct}$ are compared in Fig.~\ref{fig:tdep}.
    Twelve of the detectors showed negligible charge trapping in the string test cryostats, while only one of the BEGe detectors showed improvement with a minimal charge trapping correction. Both BEGe detectors and two of the enriched detectors required only a minimal charge trapping correction in the \MJD\ modules for all data sets, while the remaining nine enriched detectors required significant charge trapping corrections to achieve optimal energy resolution.

    The primary difference in operating conditions between the string test cryostats and the \MJD~detector modules (Fig.~\ref{fig:modules}) is the detector temperature.
    The string test cryostats were cooled by a solid copper cold finger dipped in a liquid nitrogen dewar. The modules of the \MJD, on the other hand, are cooled by a more sophisticated thermosyphon configuration~\cite{Aguayo2013a} that held the coldplate and detectors at temperatures near 80~K.
    The less efficient cooling mechanism of the string test cryostats was estimated to cool the detectors only to $\approx$95~K, roughly 15~K warmer than in the \MJD~modules. The vendor-provided cryostats used the same LN-cooled cold finger configuration as the string test cryostats.

    Drifting charges that get captured in charge traps are re-released on a time scale that varies strongly with temperature. The characteristic re-release time is distinct from the trapping time constant $\tau_{\rm ct}$ introduced earlier, although both depend on the trap depth. At high temperatures, the re-release time can be short relative to the drift time, so that trapped charges are essentially immediately released and collected in time for the energy measurement. At very low temperatures, the trapping duration can be long compared to the interval between events in the volume of the charge cloud trajectory, in which case the traps ``freeze-out'' and remain occupied, and are thus unavailable for further trapping. It is only at intermediate temperatures that the traps can hold on to the charge long enough to degrade energy, but release the charge soon enough that the trap is active again for the next charge cloud traversing its vicinity.
    
    We conclude that in the \MJD\ modules, the detector temperatures are in this intermediate range where charge trapping impacts are significant. 
    The temperature in the string test cryostats seems to have been sufficiently high to lead to prompt re-release, consistent with the lack of a need for significant charge trapping corrections in those data.

\section{Correction for $t_0$ estimation bias}
\label{sec:t0error}
    Because the $t_0$ estimation algorithms described in Sec.~\ref{sec:energy} rely on identifying a threshold crossing, the $t_0$ values obtained are biased towards later times than the true $t_0$.
    Since after the charge trapping correction the energy estimation trapezoidal filter's flat top decays with a time constant of $\tau_{\rm ct}$, the bias $\Delta t_0$ is propagated into a bias $\Delta E$ in energy measurements according to
    \begin{equation}
        \label{eq:error_prop}
        \Delta E  = \frac{\Delta t_0}{\tau_{\rm ct}} E.
    \end{equation}
    Furthermore, as the ratio of the energy to the $t_0$ threshold decreases, the size of the $\Delta t_0$ increases, resulting in energy scale non-linearity.
    By estimating the size of $\Delta t_0$ as a function of energy, one can correct for this bias and improve the linearity of the energy response.

    The size of this bias can be measured by comparing $t_0$ to the time of a pulse from a coincidence event in a second detector.
    Ideally, the second detector would have a much faster response than a PPC detector. However, for \textsc{Majorana Demonstrator} data we can still measure this bias by comparing coincidences between two PPC detectors.
    Because $\gamma$~rays emitted in a cascade will be absorbed by separate detectors within 10~ns of each other, the coincidence events they produce are effectively observed simultaneously, so any difference in $t_0$ reflects measurement bias.
    
\begin{figure}[!htbp]
        \includegraphics[width=\linewidth]{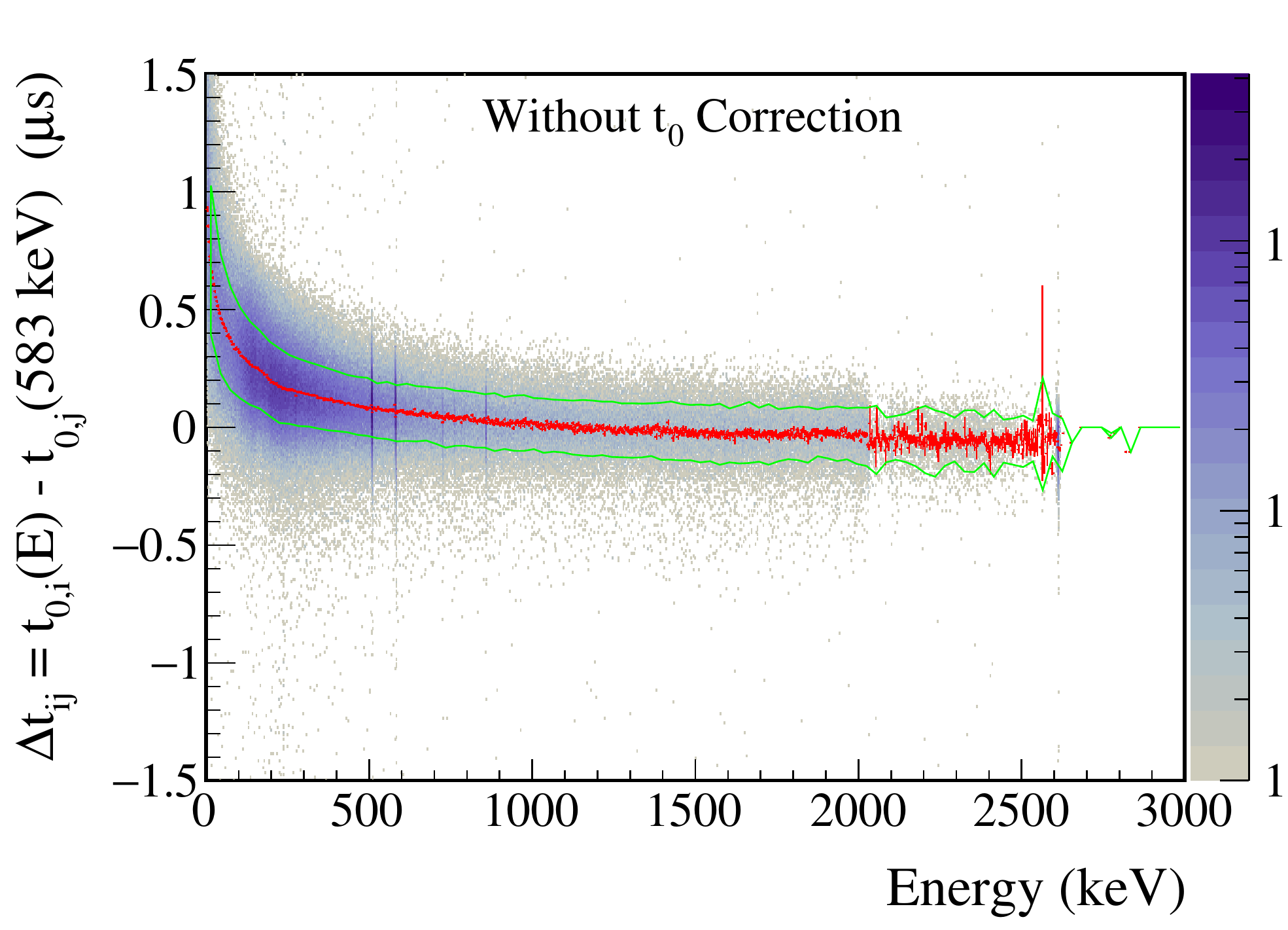}
        \includegraphics[width=\linewidth]{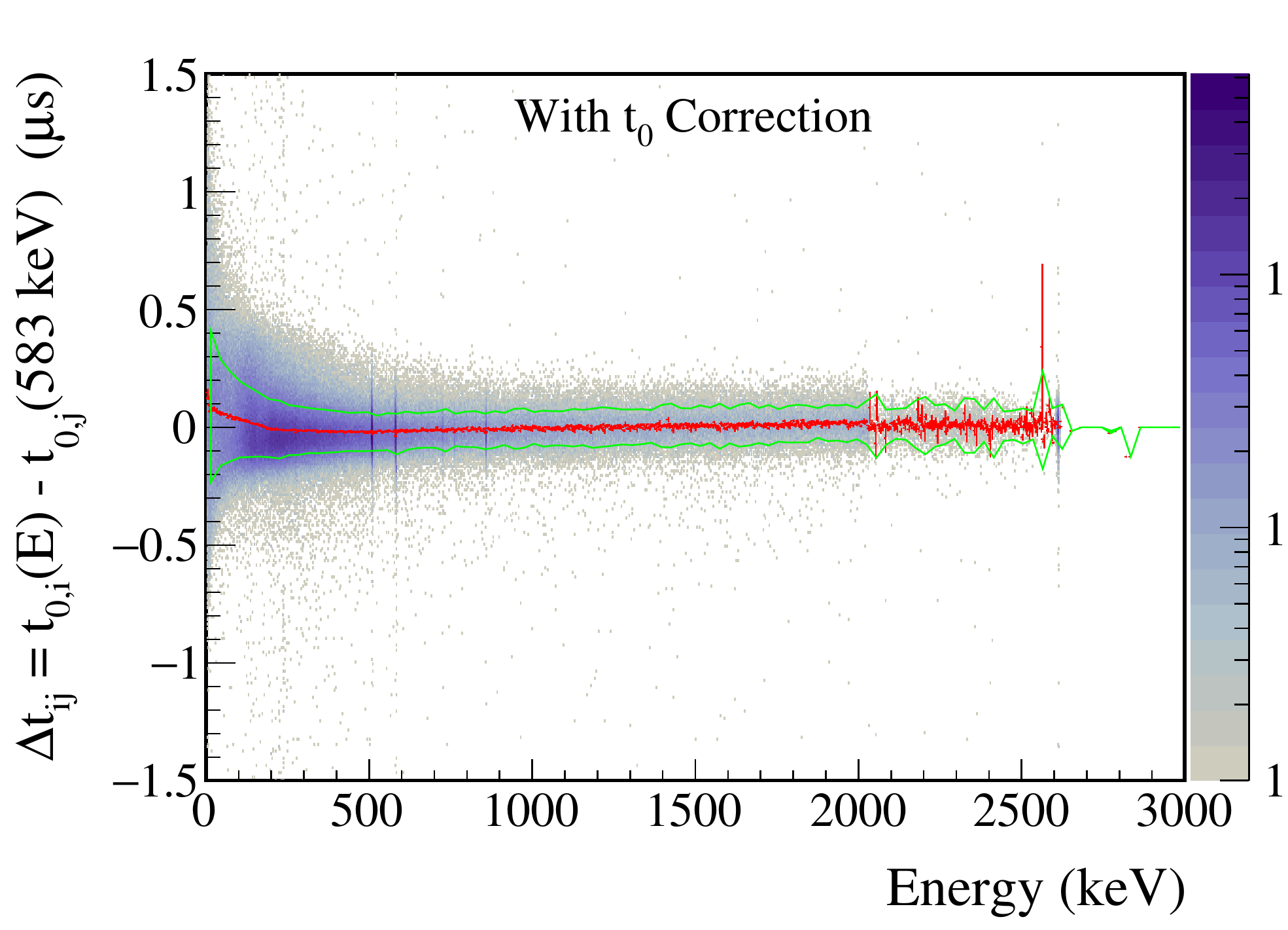}
        \caption{The timing bias calculated using Eq.~\ref{eq:t0error} between coincidence events measured during calibration of the \textsc{Majorana Demonstrator} using a $^{228}$Th line source. The coincidence events shown have one detector in the 583~keV peak, while the energy of the other detector is shown on the x-axis. The red points represent the mean bias measured for a given energy bin, relative to the 583~keV event, and the green lines show the standard deviation. On top, we see the timing bias using the $t_0$ estimate described in Sec.~\ref{sec:energy}, and on the bottom we see the bias after applying the correction described in Sec.~\ref{sec:t0error}.}
        \label{fig:t0error}
    \end{figure}

    The timing bias between detectors $i$ and $j$ is determined using
    \begin{equation}
        \label{eq:t0error}
        \Delta t_{ij} = t_{DAQ,i} + t_{0,i} - t_{DAQ,j} - t_{0,j}
    \end{equation}
    where $t_{DAQ}$ is the trigger time of the event according to the data acquisition (DAQ) system, and $t_0$ is the measured start time of the waveform relative to the first recorded sample.
    Note that $\Delta t_{ij}$ is not a clean measurement of timing bias for a single detector, but reflects the bias of all detectors involved.
    The $^{228}$Th sources used for detector calibration will emit 583~keV $\gamma$~rays followed promptly by $2615$~keV $\gamma$~rays at a high rate, which will often produce coincidence events.
    By measuring $\Delta t_{ij}$ for coincident events in which one detector measures 583~keV, we can observe the energy dependence of $\Delta t_0$ for the other detector across a wide range of energies, as shown in the top panel of Fig.~\ref{fig:t0error}. This allows us to extract a correction based on the statistics of the recorded $\Delta t_0$ values.
    
    For each detector $i$, we model its (unknown) mean $\Delta t_0$ as
    \begin{equation}
        \label{eq:t0_error_mean}
        \langle \Delta t_i(E) \rangle = \begin{cases}
            \frac{a_0}{\sqrt{E}} - a_1 & E > E_c \\
            m \cdot E + b              & E \leq E_c
        \end{cases}
    \end{equation}
    where $a_0$, $a_1$, and $E_c$ are floating model parameters, and the mean refers to the average over detector signal shapes.
    The $\frac{a_0}{\sqrt{E}}$ term can be derived by assuming that the waveform begins rising linearly from the baseline with a slope that is a constant fraction of the total energy, with a kink at the true event start time.
    The constant $a_1$ term represents a fixed offset produced by the DAQ channel.
    This model breaks down at low energies, so we transition to a linear model below energy $E_c$, with $m$ and $b$ chosen so that the function is continuous and differentiable at this transition.
    This transition typically occurs around 10~keV, below which the threshold crossing occurs after the assumption of a linear rise no longer applies.
    In addition, we model the variance of the timing bias vs energy, using
    \begin{equation}
        \label{eq:t0_error_variance}
        \langle \Delta t_i(E)^2 \rangle = \sigma_0^2 + \frac{\sigma_1^2}{E}
    \end{equation}
    where $\sigma_0$ and $\sigma_1$ are additional floating model parameters.

    Using this model, we can measure the timing bias as a function of energy for each detector by performing a simultaneous fit of the measured timing biases $\Delta t_{ij}$ of coincidence events in $^{228}$Th calibration data.
    In order to maximize the number of available events and reduce statistical error, all coincidence events are used as opposed to just those involving a 583~keV hit shown in Fig.~\ref{fig:t0error}; while not necessarily optimal, this approach works well, as seen below.
    We construct the following unbinned negative log-likelihood function:
    \begin{equation}
        \label{eq:nll}
        -2\log \mathcal{L} = \sum_{i,j} \frac{\big(\Delta t_{ij} - \langle\Delta t_i(E_i)\rangle + \langle\Delta t_j(E_j) \big\rangle)^2}{\langle \Delta t_i(E_i)^2 \rangle + \langle \Delta t_j(E_j)^2 \rangle},
    \end{equation}
    where the sum runs over all pairs of hits $i,j$ that trigger within 4~$\mu$s of each other.
    The MIGRAD algorithm from ROOT's {\sc minuit} package~\cite{BRUN1997} is used to minimize this likelihood function in order to obtain best fit parameters.
    This fit is performed on a weekly basis, for each calibration of the \textsc{Majorana Demonstrator}.
    After obtaining the best fit parameters for each channel, we correct the $t_0$ parameters for detector $i$ as follows:
    \begin{equation}
        \label{eq:t0_correction}
        t_{0,corr} = t_0 - \langle \Delta t_i(E) \rangle
    \end{equation}
    where the corrected start time $t_{0,corr}$ is then used in identifying the pickoff time for the procedure described in Sec.~\ref{sec:energy}.
    The bottom panel of Fig.~\ref{fig:t0error} shows the timing bias for coincidence events in $^{228}$Th data involving 583~keV hits after applying this correction.

\begin{figure}[!htbp]
        \includegraphics[width=\linewidth]{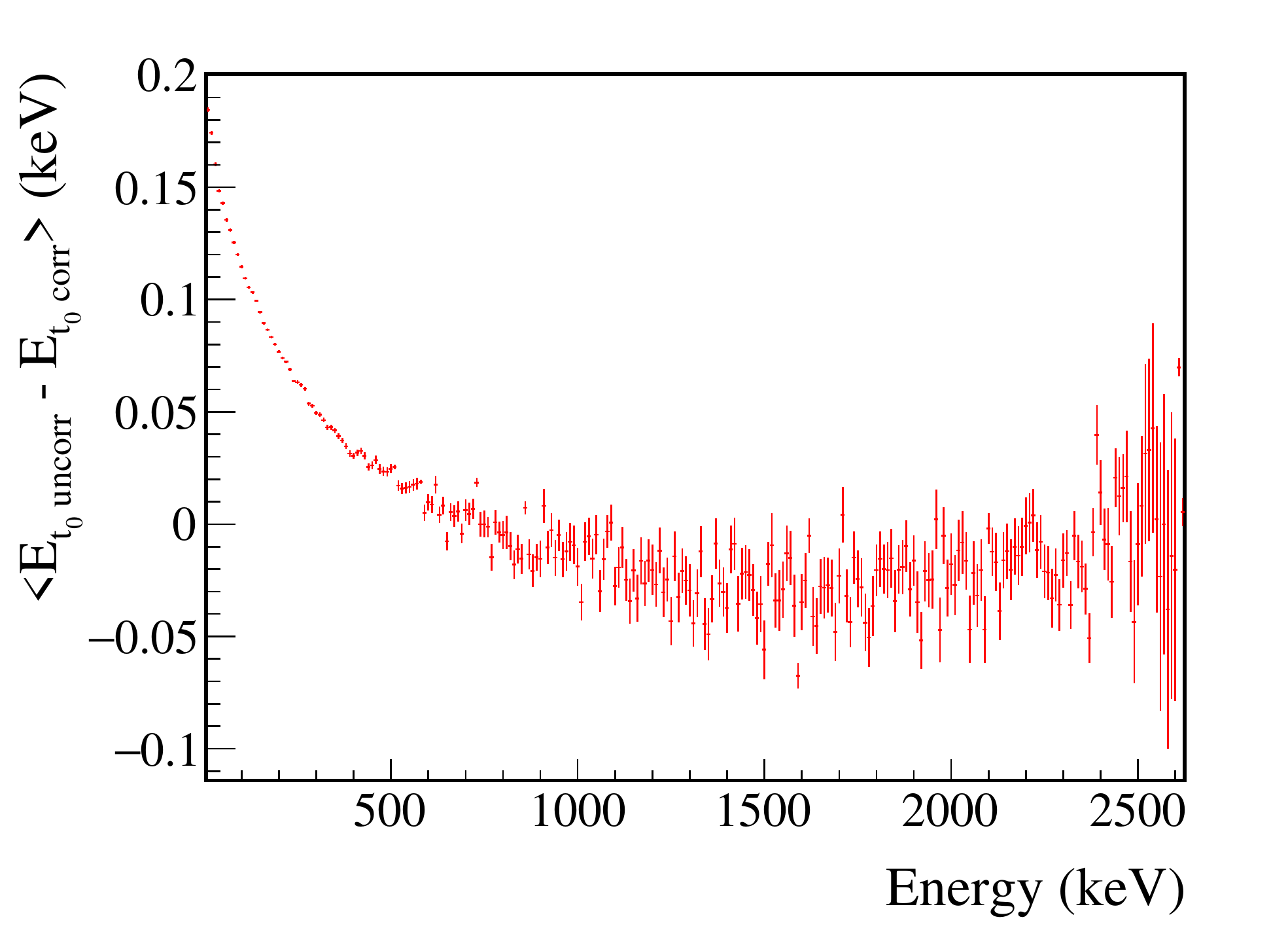}
        \includegraphics[width=\linewidth]{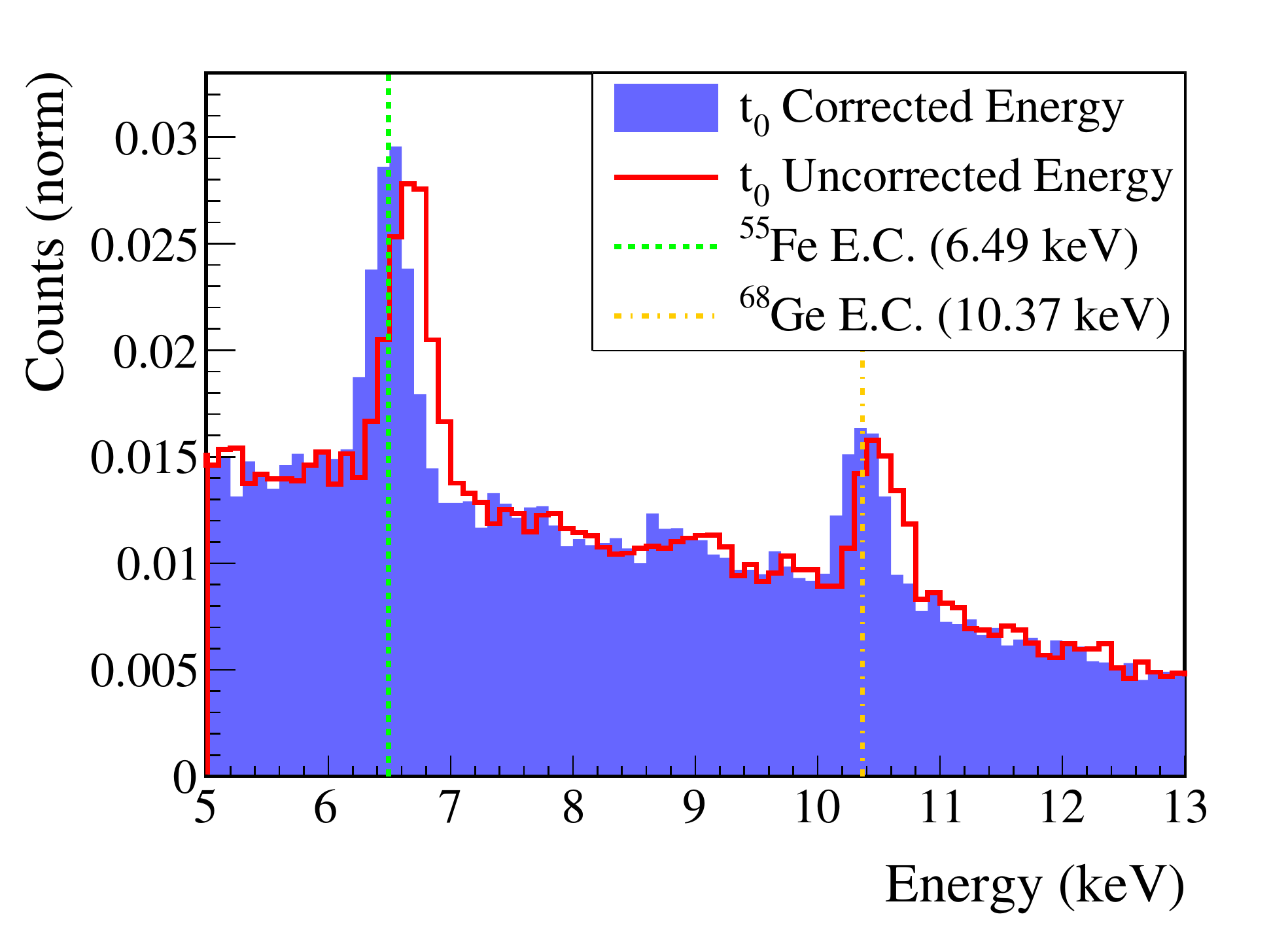}
        \caption{Top: the mean energy shift in natural isotopic abundance BEGe detectors measured during a $^{228}$Th calibration run between the energy estimator using an uncorrected $t_0$ ($E_{t_0\,\textrm{uncorr}}$) and the estimator that does perform this correction ($E_{t_0\,\textrm{corr}}$) as a function of energy. Bottom: a low energy background spectrum in natural Ge detectors, with two peaks produced by electron capture in the bulk of the detectors by cosmogenically produced isotopes $^{55}$Fe and $^{68}$Ge, with energies at 6.49~keV and 10.37~keV, respectively. Comparing the $E_{t_0\,\textrm{corr}}$ spectrum (solid blue) to the $E_{t_0\,\textrm{uncorr}}$ spectrum (red line), we see the $0.16$~keV shift produced by this correction so that the corrected peaks accurately measure the expected energies.
        \label{fig:t0error_corr_E}}
    \end{figure}

    The \textsc{Majorana Demonstrator} has performed two energy estimations which we will compare.
    The first, referred to as $t_0$ uncorrected energy, includes the charge trapping correction using a $[1, 1.5, 1]$ leading-edge trapezoidal filter and no $t_0$ correction, and is calibrated to a linear function with an offset at $E=0$~\cite{Aalseth:2017,Alvis:2019sil}.
    The second, referred to as $t_0$ corrected energy, uses the $[0.04, 0.1, 2]$ leading-edge trapezoidal filter and includes the $t_0$ correction outlined here, and is calibrated to a quadratic function with no offset at $E=0$~\cite{Arnquist:2022zrp}.
    Both estimators followed the resolution optimization procedure described in Sec.~\ref{sec:resolution}.
    Figure~\ref{fig:t0error_corr_E} shows the energy shift between these estimators.
    Prior to $t_0$ correction, the energy bias is greatest at low energies, with an average value of $\approx0.15$~keV, and reaching up to $0.3$~keV in detectors with the largest effect, which is equivalent to $\approx$1~FWHM of a low energy peak.
    By correcting this nonlinearity, we can achieve accurate energy measurements at all energies, and we remove the need for a calibration offset at $E=0$.
    This also removes the need for a separate calibration at low energies as was performed in Ref.~\cite{Vorren:2017}, and aids in analyses that utilize events at all energies.

\section{An Alternative Correction}
\label{sec:alternative}
    An alternative charge trapping algorithm was developed in parallel with the $\tau$ optimization algorithm described above and implemented in a secondary, redundant analysis.
    The alternative method was tested using \MJD\ calibration data and gives results identical to the method described in Sec.~\ref{sec:energy} within statistical uncertainty, although the alternative method was found to sometimes out perform the $\tau$ optimization method.

\begin{figure}[!htbp]
        \centering
        \includegraphics[width=\columnwidth]{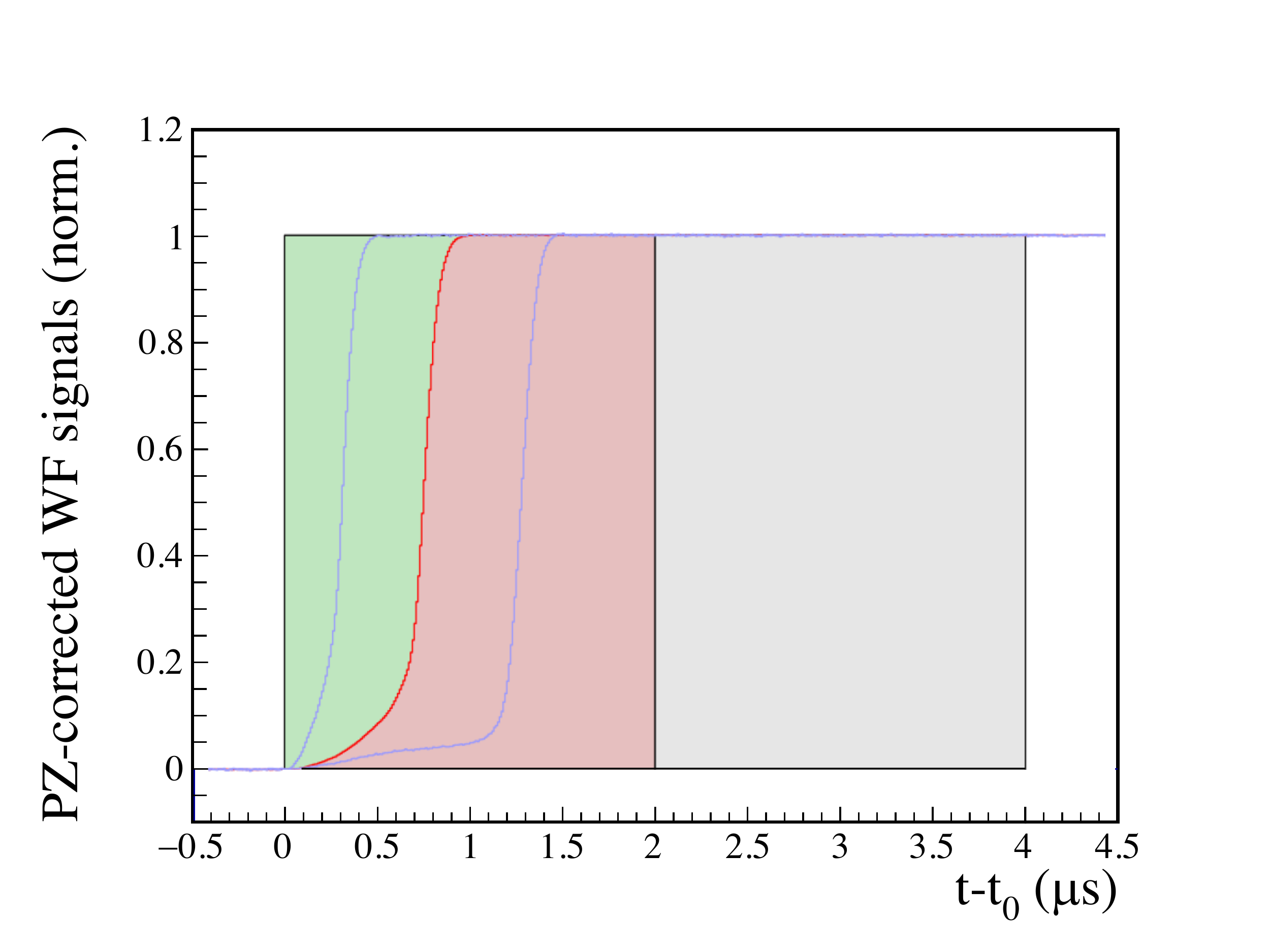}
        \caption{Visualization of the alternative charge trapping correction. The value of $D$ is represented by the area shaded green, given by the difference between the 2-$\mu$s integrated waveform regions represented by gray and red. For reference, pulses with shorter and longer drift times are included; these would have smaller and larger values of $D$, respectively.}
        \label{fig:ctc2}
    \end{figure}

    In the alternative method, a standard pole-zero correction is applied using the measured preamplifier time constant only, $\tau_{RC}$.
    This generates outputs similar to the blue signals in Fig.~\ref{fig:ctsim}.
    The start time, $t_0$, is evaluated using the same method described in Sec.~\ref{sec:energy}. Then the pole-zero-corrected waveform signal is integrated over two time intervals, [$t_0 , t_0+2\mu s$] and [$t_0 + 2\mu s, t_0+4\mu s$], and the difference $D$ between the two integrals is calculated.
    The 2 $\mu s$ integration time is chosen because it is slightly longer than the maximum drift time for all of the \MJD~detectors. 
    The integration windows and the calculation of $D$ are illustrated in Fig.~\ref{fig:ctc2}. 

    As described in more detail in the Appendix, the signal degradation due to charge trapping is nearly proportional to the integral of the uncollected charge,  which is approximately proportional to $D$. 
    Thus the final step is to add a small $D$-dependent correction to the uncalibrated energy calculated using the standard trapezoidal filter. In the \MJD\ PPCs a simple linear correction proved to be sufficient:
    \begin{equation}
        E_\mathrm{corrected} = E_\mathrm{uncorrected} + \alpha D.
        \label{eq:EcorrAlt}
    \end{equation}
    The charge-trapping correction factor $\alpha$ is optimized by minimizing the FWHM of the 2615-keV peak.

    This alternative algorithm has several advantages. The primary advantage is that optimization of $\alpha$ is much less computationally intensive than optimization of $\tau$, which requires repeated digital signal processing.
    This algorithm should also have some advantage at higher count rates because it does not generate any undershoot at the end of the trapezoidal filter that results from the incomplete pole-zero correction of the $\tau$ optimization method. 
    This algorithm is also superior to methods based on estimating the drift time from the pulse rise time because the evaluation of $D$ is based on an integral and requires the estimation of only one time point ($t_0$) rather than two. Nonlinear dependence of the correction on $D$ can be easily accommodated by extending Eq.~\ref{eq:EcorrAlt} to quadratic or higher order. $D$ is also useful in contexts outside of energy estimation, as is the pole-zero corrected waveform based on $\tau_{RC}$ only.
    
    This alternative method and the original $\tau$ optimization method assume different time dependences for the untrapped charges.
    The $\tau$ optimization method assumes exponential decay of untrapped charge during the charge drift, while the alternative method assumes a linear decay.
    This linear approximation is valid as long as the charge trapping rate is small, i.e., if the maximum drift time is short compared to the mean free drift time.
    It would be possible to extend the alternative method to accommodate exponential decay by modifying Eq.~\ref{eq:EcorrAlt}.
    Both methods require an accurate determination of $t_0$ for each signal, but the bias correction described in Sec.~\ref{sec:t0error} works equally well for both.
    Finally, both algorithms assume that in a PPC detector it is only hole trapping that significantly affects the extracted energy, and that the interaction creating the electron-hole pairs is away from the point contact, i.e.~at a small weighting potential.
    Both of these assumptions are valid for a majority of signals in PPC detectors.
    
\section{Conclusion}
\label{sec:conclusion}
    We reported two charge trapping corrections
    that result in a significant improvement in the energy performance of the \MJD. One is
    based on a modified pole-zero-corrected energy estimator, the other is based on an integral of the uncollected charge.
    An analysis based on the former gives an average FWHM after charge trapping corrections of 2.4 keV at the \nonubb~$Q$ value of 2039 keV and 2.9 keV at 2615 keV (statistical uncertainties only).
    The presence or absence of significant charge trapping appears to strongly correlate with detector temperature, requiring energy estimation parameters to be tuned to cryostat-specific environmental conditions. These correction algorithms are very flexible and suitable for other arrays of HPGe PPC detectors, including the upcoming LEGEND experiment~\cite{Abgrall:2017syy,LEGEND:2021bnm}.

\section*{Acknowledgments}
This material is based upon work supported by the U.S.~Department of Energy, Office of Science, Office of Nuclear Physics under contract/award nos.~DE-AC02-05CH11231, DE-AC05-00OR22725, DE-AC05-76RL0130, DE-FG02-97ER41020, DE-FG02-97ER41033, DE-FG02-97ER41041, DE-SC0012612, DE-SC0014445, DE-SC0018060, DE-SC0022339, and LANLEM77/LANLEM78. We acknowledge support from the Particle Astrophysics Program and Nuclear Physics Program of the National Science Foundation through grant nos.~MRI-0923142, PHY-1003399, PHY-1102292, PHY-1206314, PHY-1614611, PHY-1812409, PHY-1812356, PHY-2111140, and PHY-2209530. We gratefully acknowledge the support of the Laboratory Directed Research \& Development (LDRD) program at Lawrence Berkeley National Laboratory for this work. We gratefully acknowledge the support of the U.S.~Department of Energy through the Los Alamos National Laboratory LDRD Program and through the Pacific Northwest National Laboratory LDRD Program for this work.  We gratefully acknowledge the support of the South Dakota Board of Regents Competitive Research Grant.
We acknowledge the support of the Natural Sciences and Engineering Research Council of Canada, funding reference number SAPIN-2017-00023, and from the Canada Foundation for Innovation John R.~Evans Leaders Fund.  This research used resources provided by the Oak Ridge Leadership Computing Facility at Oak Ridge National Laboratory and by the National Energy Research Scientific Computing Center, a U.S.~Department of Energy Office of Science User Facility. We thank our hosts and colleagues at the Sanford Underground Research Facility for their support.
    
\appendix

\section{Signal Induction with Charge Trapping}
\label{app:signal}

Signal induction in PPC HPGe detectors is governed by the Shockley-Ramo theorem~\cite{shockley1938,Ramo:1939vr}: the induced current $I$ at a conductor due to a charge $q$ moving with velocity $\vec{v}$ in its vicinity is given by
\begin{equation}
I(t) = q \, \vec{v} \cdot \vec{\nabla} \phi,
\end{equation}
where $\phi$ is the weighting potential for the conductor, evaluated by giving the conductor unit potential and grounding all other conductors in the vicinity.

In PPC HPGe detectors, the signal at the point contact is induced by the motion of both electron ($e$) and hole ($h$) charge clouds. The holes drift toward the point contact while electrons drift away from it, with differing drift speeds. In the presence of charge trapping, the charge $q$ in each cloud decreases with time. Assuming a constant charge trapping probability that does not vary with position, the charge in each cloud will decay exponentially, with species dependent time constants. Ignoring phenomena like recombination, diffusion, and space charge effects, the induced current at the point contact can be written
\begin{equation}
I(t) = q_0 e^{-(t-t_0)/\tau^h_{\rm ct}} \frac{d\phi_h}{dt} - q_0 e^{-(t-t_0)/\tau^e_{\rm ct}} \frac{d\phi_e}{dt},
\label{eq:Iinduced}
\end{equation}
where $q_0$ is the initial charge in the hole cloud at time $t_0$, $\tau^k_{\rm ct}$ is the charge trapping time constant for species $k$, and $\frac{d\phi_k}{dt}$ is the rate of change of the weighting potential traversed by species $k$. $\frac{d\phi_k}{dt}$ depends not only on the charge species but also the trajectory, and is non-zero only for $t_0 < t < t_0 + T_k$,  where $T_k$ is the time it takes the charge cloud to reach a detector contact.

This signal is measured by collecting the induced current on a capacitor and measuring the charge $Q(t)$, given by the integral of Eq.~\ref{eq:Iinduced}. The electrons and holes start at the same location with the same initial weighting potential $\phi_0$. For negligible charge trapping ($\tau^k_{\rm ct} \gg T_k$), the collected charge is given by
\begin{equation}
    Q_{\rm no~ct}(t) = q_0 \Delta \phi(t)
    \label{eq:Qnoct}
\end{equation}
where $\Delta \phi(t)$ is the difference in weighting potential at the positions of the hole and electron clouds at time $t$. This difference is zero for $t < t_0$, and it is equal to 1 for $t > T_e, T_h$.

\subsection{Effective Pole-Zero Correction}
\label{sec:effPZmath}

The effective pole-zero correction described in Sec.~\ref{sec:trapping} can be thought of as first deconvolving the electronic response from the recorded signals, and then convolving the signal with an exponential response with optimized time constant $\tau_{\rm ct}$. The resulting signal has modified collected charge
\begin{equation}
    \tilde{Q}(t) = \int_{t_0}^t e^{-\frac{t-t'}{\tau_{\rm ct}}} \, I(t') \, dt'.
\end{equation}
The term $e^{t'/\tau_{\rm ct}}$ can be combined with the $e^{-t'/\tau^k_{\rm ct}}$ terms in $I(t')$ that represent the charge loss during drift due to trapping. Thus the convolution has the effect of ``adding back'' the lost charge. In PPC detectors the signal generation for most events is dominated by the hole motion, so the optimal value of $\tau_{\rm ct}$ will be close to $\tau^h_{\rm ct}$. Thus hole trapping is almost fully compensated, while the electron trapping, which is more rapid, remains only partially compensated but contributes less to signal degradation. The resulting modified collected charge is thus approximately
\begin{equation}
    \tilde{Q}(t) \approx e^{-(t-t_0)/\tau_{\rm ct}} \, q_0 \, \Delta \phi(t), 
\end{equation}
which differs from Eq.~\ref{eq:Qnoct} by only a multiplicative exponential function of time. As a result, the fixed-time pickoff energy estimation method described in Sec.~\ref{sec:energy} will obtain a similar value for events with the same energy deposition at different points in the detector, with minimal variation due to charge trapping.

\subsection{Uncollected Charge Integral Method}

The alternative approach described in Sec.~\ref{sec:alternative} can be motivated by approximating the exponential functions in Eq.~\ref{eq:Iinduced} by their first-order Taylor expansions and then directly integrating.  
The terms like $t \frac{d\phi}{dt}$ can be integrated by parts to give
\begin{equation}
    \int_0^T t \, \frac{df}{dt} \, dt = \int_0^T [f(T) - f(t)] \, dt.
\end{equation}
Using the fact that $\phi_h(t > t_0 + T_h) = 1$ and $\phi_e(t > t_0 + T_e) = 0$, we find that at the end of the drift the collected charge has approximately the constant value:
\begin{equation}
    Q \approx q_0 - \frac{q_0}{\tau_{\rm ct}^h} \int_{t_0}^t \left(1 - \Delta\phi(t') + \frac{\tau_{\rm ct}^h-\tau_{\rm ct}^e}{\tau_{\rm ct}^e} \phi_e(t') \right) \,dt', 
    \label{eq:Qcoll}
\end{equation}
where $t$ is any time satisfying $t - t_0 > T_h, T_e$.

The quantity $q_0 (1- \Delta \phi(t))$ represents the uncollected charge at time $t$ (in the absence of charge trapping). To zeroth order in $t/\tau^h_{\rm CT}$, its integral over the drift is proportional to the quantity $D$ defined in Sec.~\ref{sec:alternative}. The third term in Eq.~\ref{eq:Qcoll} gives a contribution proportional to the integral of the signal due to the electron drift only. The integrated electron signal is small for the vast majority of events, and is comparable to the uncollected charge integral only for events originating very close to the point contact. On average it contributes a shift that is proportional to energy, and hence is accounted for in the energy scale calibration.
Thus, an energy correction proportional to $D$ is expected to remove most of the variance due to charge trapping.

\bibliography{reference}
\end{document}